\documentclass[AMA,STIX1COL]{WileyNJD-v2}
\usepackage{moreverb}

\usepackage{url}
\usepackage{graphicx}
\usepackage{amsmath} 
\usepackage{multirow}
\usepackage{anyfontsize}
\allowdisplaybreaks
\newcommand{\I}{\boldsymbol{I}}
\newcommand{\J}{\boldsymbol{J}}

\newcommand{\X}{\boldsymbol{X}}
\newcommand{\Z}{\boldsymbol{Z}}
\newcommand{\V}{\boldsymbol{V}}

\newcommand{\Y}{\boldsymbol{Y}}
\newcommand{\bb}{\boldsymbol{b}}
\newcommand{\s}{\boldsymbol{s}}
\newcommand{\W}{\boldsymbol{W}}
\newcommand{\D}{\boldsymbol{D}}
\newcommand{\M}{\boldsymbol{M}}
\newcommand{\bc}{\boldsymbol{c}}
\newcommand{\0}{\boldsymbol{0}}
\newcommand{\btheta}{\boldsymbol{\theta}}
\newcommand{\bdelta}{\boldsymbol{\delta}}
\newcommand{\bOmega}{\boldsymbol{\Omega}}
\newcommand{\bLambda}{\boldsymbol{\Lambda}}
\newcommand{\bbeta}{\boldsymbol{\beta}}
\newcommand{\bbbeta}{\boldsymbol{\eta}}
\newcommand{\bepsilon}{\boldsymbol{\epsilon}}
\newcommand{\bgamma}{\boldsymbol{\gamma}}
\newcommand{\bGamma}{\boldsymbol{\Gamma}}
\newcommand{\bphi}{\boldsymbol{\phi}}
\newcommand{\bsigma}{\boldsymbol{\sigma}}
\newcommand{\bSigma}{\boldsymbol{\Sigma}}

\newcommand{\bomega}{\boldsymbol{\omega}}
\DeclareMathOperator{\Exp}{\mathbb{E}}
\DeclareMathOperator{\Ind}{\mathbb{I}}

\newcommand\BibTeX{{\rmfamily B\kern-.05em \textsc{i\kern-.025em b}\kern-.08em
T\kern-.1667em\lower.7ex\hbox{E}\kern-.125emX}}

\articletype{Research Article}%

\received{<day> <Month>, <year>}
\revised{<day> <Month>, <year>}
\accepted{<day> <Month>, <year>}

\begin{document}

\title{Power analyses for stepped wedge designs with multivariate continuous outcomes}

\author[1,2,3]{Kendra Davis-Plourde*}

\author[4,5]{Monica Taljaard}

\author[1,3]{Fan Li**}

\authormark{DAVIS-PLOURDE \textsc{et al}}

\address[1]{\orgdiv{Department of Biostatistics}, \orgname{Yale School of Public Health}, \orgaddress{\city{New Haven}, \state{Connecticut}, \country{U.S.A.}}}

\address[2]{\orgdiv{Department of Internal Medicine}, \orgname{Yale School of Medicine}, \orgaddress{\city{New Haven}, \state{Connecticut}, \country{U.S.A.}}}

\address[3]{\orgdiv{Center for Methods in Implementation and Prevention Science}, \orgname{Yale School of Public Health}, \orgaddress{\city{New Haven}, \state{Connecticut}, \country{U.S.A.}}}

\address[4]{\orgdiv{Clinical Epidemiology Program}, \orgname{Ottawa Hospital Research Institute}, \orgaddress{\state{Ottawa}, \country{Canada}}}

\address[5]{\orgdiv{School of Epidemiology and Public Heath}, \orgname{University of Ottawa}, \orgaddress{\state{Ottawa}, \country{Canada}}}

\corres{*Kendra Davis-Plourde, Department of Biostatistics, Yale School of Public Health, New Haven, CT 06520.\\ \email{kendra.plourde@yale.edu}\\\\
**Fan Li, Department of Biostatistics, Yale School of Public Health, New Haven, CT 06520. \email{fan.f.li@yale.edu}}


\abstract[Abstract]{Multivariate outcomes are common in pragmatic cluster randomized trials. While sample size calculation procedures for multivariate outcomes exist under parallel assignment, none have been developed for a stepped wedge design. In this article, we present computationally efficient power and sample size procedures for stepped wedge cluster randomized trials (SW-CRTs) with multivariate outcomes that differentiate the within-period and between-period intracluster correlation coefficients (ICCs). Under a multivariate linear mixed model, we derive the joint distribution of the intervention test statistics which can be used for determining power under different hypotheses and provide an example using the commonly utilized intersection-union test for co-primary outcomes. Simplifications under a common treatment effect and common ICCs across endpoints and an extension to closed cohort designs are also provided. Finally, under the common ICC across endpoints assumption, we formally prove that the multivariate linear mixed model leads to a more efficient treatment effect estimator compared to the univariate linear mixed model, providing a rigorous justification on the use of the former with multivariate outcomes. We illustrate application of the proposed methods using data from an existing SW-CRT and present extensive simulations to validate the methods.}



\maketitle


\section{Introduction}\label{sec:intro}

\begin{figure}
\setlength{\unitlength}{0.14in} 
\centering
\begin{picture}(26,17) 
\setlength\fboxsep{0pt}
\put(1,14){\framebox(4,2.5){}}
\put(7,14){\colorbox{gray!40}{\framebox(4,2.5){}}}
\put(13,14){\colorbox{gray!40}{\framebox(4,2.5){}}}
\put(19,14){\colorbox{gray!40}{\framebox(4,2.5){}}}
\put(25,14){\colorbox{gray!40}{\framebox(4,2.5){}}}
\put(1,10){\framebox(4,2.5){}}
\put(7,10){\framebox(4,2.5){}}
\put(13,10){\colorbox{gray!40}{\framebox(4,2.5){}}}
\put(19,10){\colorbox{gray!40}{\framebox(4,2.5){}}}
\put(25,10){\colorbox{gray!40}{\framebox(4,2.5){}}}
\put(1,6){\framebox(4,2.5){}}
\put(7,6){\framebox(4,2.5){}}
\put(13,6){\framebox(4,2.5){}}
\put(19,6){\colorbox{gray!40}{\framebox(4,2.5){}}}
\put(25,6){\colorbox{gray!40}{\framebox(4,2.5){}}}
\put(1,2){\framebox(4,2.5){}}
\put(7,2){\framebox(4,2.5){}}
\put(13,2){\framebox(4,2.5){}}
\put(19,2){\framebox(4,2.5){}}
\put(25,2){\colorbox{gray!40}{\framebox(4,2.5){}}}
\put(-5,3.1){\large Sequence $4$}
\put(-5,7.1){\large Sequence $3$}
\put(-5,11.1){\large Sequence $2$}
\put(-5,15.1){\large Sequence $1$}
\put(1.3,17){\large Period $1$}\put(7.3,17){\large Period $2$}
\put(13.3,17){\large Period $3$}\put(19.3,17){\large Period $4$}\put(25.3,17){\large Period $5$}
\end{picture}
\caption{A schematic illustration of a stepped wedge cluster randomized trial (SW-CRT) with five periods and four distinct intervention sequences. Clusters are randomized to a sequence and typically each sequence contains the same number of clusters. Each white cell indicates the control condition and each gray cell indicates the intervention condition.} 
\label{f:swcrt_ex} 
\end{figure}
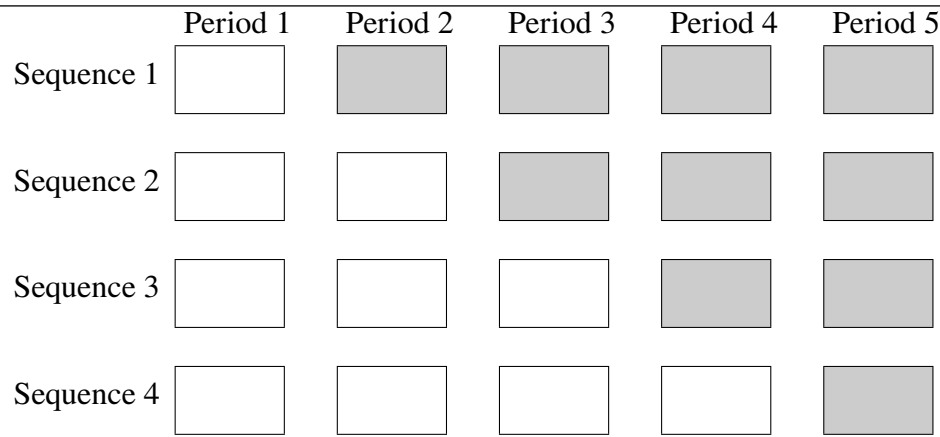

The stepped wedge cluster randomized trial (SW-CRT) is an increasingly popular design, typically used to evaluate health system, policy and service delivery interventions in real-world settings.  In this design, clusters typically all start in the control condition; at different intervals, clusters are randomly selected to cross over to the intervention condition until by the end of the trial, all clusters are in the intervention condition. Clusters that share the same cross over point are said to be a part of the same treatment sequence (Figure \ref{f:swcrt_ex}). This design is often chosen because it allows all clusters to receive the intervention during the trial and for its potential gains in statistical efficiency.\citep{hemming2020reflection} At each interval, measurements are taken in each cluster: in a cross-sectional design, the measurements are taken on different individuals, and in a closed-cohort design, on the same individuals.\citep{copas2015designing}

It is well-known that, in a cluster randomized design, individual outcomes within a cluster are more similar than those between clusters. Thus, rigorous design and analysis require statistical methods that take into account the correlation within clusters: referred to as the intracluster correlation coefficient (ICC). SW-CRTs have an additional layer of complexity due to its longitudinal nature: individual measurements within the same cluster and the same time period can be expected to be more strongly correlated than measurements within the same cluster but different time periods. Thus, the design and analysis of SW-CRTs commonly include a within-period ICC and a between-period ICC. Under a cross-sectional design, a typical analytical model for SW-CRTs is the Hooper et al.\citep{hooper2016sample} and Girling et al.\citep{Girling2016} linear mixed effects model (LMM) which includes a random cluster effect, allowing individuals within the same cluster to be correlated, and a random cluster-by-period interaction, allowing individuals within the same cluster at the same time period to share an additional level of correlation. The two random effects are assumed to be independent and their inclusion in the LMM induces the so-called nested exchangeable correlation structure.\citep{li2021mixed} Further, allowance for the random cluster-by-period effect has been shown to be vital in the design phase of SW-CRTs to avoid underpowered trials.\citep{taljaard2016substantial}

While most trials have a single primary outcome which drives the sample size calculation, multivariate or co-primary outcomes are increasingly common in trials designed with a pragmatic intention, for a variety of reasons.\citep{taljaard2021methodological} For example, multiple co-primary outcomes may be selected to satisfy the decision-making needs of a range of trial stakeholders and to ensure that the trial is adequately powered to detect effects on both clinical and patient-reported outcomes. Multivariate outcomes are also common when the intervention  specifically targets multiple types of participants, for example, the patient-caregiver dyad in trials involving the elderly. Finally, patient-oriented outcomes may be measured using questionnaires with multiple subscales and the intervention effects on each subscale may be of interest. 

Guidance for the design and analysis of CRTs with multivariate outcomes is currently available only for parallel-arm designs. Turner et al.\citep{turner2006modelling} proposed a multivariate linear mixed model (MLMM) for analyzing multivariate normally distributed outcomes, where between-outcome correlations on the cluster and individual levels are taken into account via random effects. Based on this MLMM, Yang et al.\citep{yang2022power} recently developed sample size considerations for parallel-arm CRTs with multivariate outcomes, using an intersection-union test (IU-test) \citep{sozu2010sample,sozu2012sample} which requires statistical significance on all outcomes, i.e. a co-primary outcome approach. To the best of our knowledge, there are currently no available methods for designing SW-CRTs with multivariate outcomes. Sample size and power considerations are more complicated in SW-CRTs because of the need to account for both within-period and between-period ICCs. 

Our study is motivated by the shared decision-making in inter-professional home care teams (IP-SDM) study. The IP-SDM study is a cross-sectional SW-CRT evaluating the implementation of shared decision-making tools in inter-professional home care teams caring for elderly clients and their caregivers.\citep{adisso2022shared} The primary outcome of interest was the binary decision to stay at home or move while a key secondary outcome was quality of life measured using two subscales of the Nottingham Health Profile (NHP): social isolation and emotional reactions \citep{zengin2014assessment} used to assess social and personal effects of illness in patients. Although the NHP subscales were considered secondary outcomes in the IP-SDM study, here we illustrate how a future trial might be designed with the NHP subscales as the primary focus. Designing SW-CRTs with more than one primary outcome, in fact, has become increasingly common in trials in the frail elderly. For example, the Connect-Home study is a SW-CRT to evaluate a transitional care intervention in seriously ill patients being discharged from skilled nursing facilities to their caregivers.\citep{toles2021transitional} The co-primary outcomes were the preparedness of patients for discharge and the preparedness of caregivers for providing care, both measured using questionnaires. Another recent example is a SW-CRT examining the effect of an interdisciplinary medication review on quality of life measures within older populations with polypharmacy.\citep{bosch2021effect} The investigators had two primary quality of life measures of interest, one of which included eight subscales which were each modeled separately. In individually randomized trials, ignoring the potential correlations between multiple primary endpoints (referred to as the between-endpoint ICC) can often lead to larger than necessary sample size estimates.\citep{Micheaux2014power} In SW-CRTs, the impact of the between-endpoint ICC may be more complex since outcomes are collected over multiple periods and the within-period and between-period ICCs need to be considered. 

In this article, we develop novel design formulas to enable computationally efficient power and sample size calculation for designing cross-sectional SW-CRTs with multivariate outcomes that differentiate the within-period and between-period ICCs. In Section \ref{s:model}, we provide an extension of the standard LMM for cross-sectional SW-CRTs with a single outcome to a MLMM where the effects of the intervention on multiple outcomes are simultaneously estimated and evaluated. In Section \ref{s:power}, we derive the joint distribution of the intervention test statistics which can be used for generating power estimates under any specified hypothesis and provide an example using the commonly utilized IU-test for co-primary outcomes. Further, we provide insights into power calculations in the special case of assuming a common treatment effect across outcomes, assuming common ICCs across outcomes, and under closed-cohort designs. In Section \ref{s:application}, we illustrate our sample size methodology using the quality of life measures from the IP-SDM study, and we further validate our power procedure under the IU-test using simulations in Section \ref{s:simulation}. Finally, we conclude with a discussion in Section \ref{s:discussion}.


\section{Modelling Multivariate Continuous Outcomes in SW-CRTs}
\label{s:model}

We first consider a cross-sectional SW-CRT with $L$ continuous outcomes, and assume that the scientific interest lies in simultaneously estimating the effect of an intervention on each outcome. In this article, we propose a multivariate version of the commonly used LMM \citep{hooper2016sample,Girling2016} for analyzing cross-sectional SW-CRTs with a single outcome. Specifically, let $\Y_{ijk}=(Y_{ijk1},\dots,Y_{ijkL})^\top$ be the vector of $L$ outcomes for subject $k=\{1,\ldots,N_{ij}\}$ nested in cluster $i=\{1,\ldots,I\}$ at time period $j=\{1,\ldots,T\}$, we simultaneously model these outcomes using the following MLMM
\begin{align}\label{eq:MLMM}
    \Y_{ijk}&=\bbeta_{0}+\bbeta_{j}+X_{ij}\bdelta+\bb_{i}+\s_{ij}+\bepsilon_{ijk},
\end{align}
where $X_{ij}$ is an indicator variable for the intervention (equal to 1 if cluster $i$ is receiving the intervention at time period $j$ and 0 otherwise) and $\bdelta=(\delta_1,\ldots,\delta_L)^\top$ is a vector of endpoint-specific intervention effects. Under model \eqref{eq:MLMM}, $\bbeta_0=(\beta_{01},\ldots,\beta_{0L})^\top$ is a vector of endpoint-specific means under the control condition, $\bbeta_j=(\beta_{j1},\ldots,\beta_{jL})^\top$ is a vector of fixed time effects (to control for confounding by time) for period $j$ and is typically treated as a categorical variable for maximum flexibility; in other words, the above MLMM adjusts for possibly unique secular trends for each outcome $l \in \{1,\ldots,L\}$. In addition, $\bb_i=(b_{i1},\ldots,b_{iL})^\top$ is a vector of cluster random effects and follows a multivariate normal distribution denoted by $\mathcal{N}(\0_{L \times 1}, \bSigma_b)$, $\s_{ij}=(s_{ij1},\ldots,s_{ijL})^\top$ is a vector of random cluster-by-period effects and follows a multivariate normal distribution denoted by $\mathcal{N}(\0_{L\times 1},\bSigma_s)$, and $\bepsilon_{ijk}=(\epsilon_{ijk1},\ldots,\epsilon_{ijkL})^\top$ is a vector of random errors and follows a multivariate normal distribution denoted by $\mathcal{N}(\0_{L\times 1},\bSigma_\epsilon)$. We assume $\bb_i$, $\s_{ij}$, and $\bepsilon_{ijk}$ are independent for identifiability and place no further restrictions on $\bSigma_b$, $\bSigma_s$, and $\bSigma_\epsilon$ except that they are positive definite. We denote the diagonal elements of $\bSigma_b$ as $\sigma_{bl}^2$ and off-diagonal elements as $\sigma_{bll'}$ giving us a total of $L(L+1)/2$ variance components for defining $\bSigma_b$. Similarly, the diagonal elements of $\bSigma_s$ and $\bSigma_\epsilon$ are denoted by $\sigma_{sl}^2$ and $\sigma_{\epsilon l}^2$ and off-diagonal as $\sigma_{sll'}$ and $\sigma_{\epsilon ll'}$ respectively, giving $\bSigma_s$ and $\bSigma_\epsilon$ each $L(L+1)/2$ variance components. 

Under this parameterization, the marginal variance of each endpoint can vary with $l$ and is given by $\sigma^2_{yl}=\sigma^2_{bl}+\sigma^2_{sl}+\sigma^2_{\epsilon l}$. Furthermore, our model implicitly provides the following five ICCs also shown in Table \ref{t:ICCs}: 
\begin{enumerate}
    \item[(1)] $\rho_0^l=\text{corr}(Y_{ijkl},Y_{ijk'l})=(\sigma^2_{bl}+\sigma^2_{sl})/\sigma^2_{yl}$, denoting the within-period inter-subject correlation of the outcomes corresponding to the same endpoint or otherwise known as the within-period endpoint-specific ICC;
    \item[(2)] $\rho_1^l=\text{corr}(Y_{ijkl},Y_{ij'k'l})=\sigma^2_{bl}/\sigma^2_{yl}$, denoting the between-period endpoint-specific ICC; 
    \item[(3)] $\rho_0^{ll'}=\text{corr}(Y_{ijkl},Y_{ijk'l'})=(\sigma_{bll'}+\sigma_{sll'})/(\sigma_{yl}\sigma_{yl'})$, denoting the within-period inter-subject correlation of two outcomes corresponding to two different endpoints $l$ and $l'$ or otherwise known as the within-period between-endpoint ICC; 
    \item[(4)] $\rho_1^{ll'}=\text{corr}(Y_{ijkl},Y_{ij'k'l'})=\sigma_{bll'}/(\sigma_{yl}\sigma_{yl'})$, denoting the between-period between-endpoint ICC; 
    \item[(5)] $\rho_2^{ll'}=\text{corr}(Y_{ijkl},Y_{ijkl'})=(\sigma_{bll'}+\sigma_{sll'}+\sigma_{\epsilon ll'})/(\sigma_{yl}\sigma_{yl'})$, denoting the intra-subject between-endpoint ICC or for brevity, the intra-subject ICC.
\end{enumerate}
Under our model we have symmetry in that $\rho_0^{ll'}=\rho_0^{l'l}$, $\rho_1^{ll'}=\rho_1^{l'l}$, and $\rho_2^{ll'}=\rho_2^{l'l}$, and degeneracy such that $\rho_0^{ll}=\rho_0^l$, $\rho_1^{ll}=\rho_1^l$, and $\rho_2^{ll}=1$. Our ICC definitions from the MLMM \eqref{eq:MLMM} also implicitly assume $\rho_1^l \leq \rho_0^l$ and $\rho_1^{ll'} \leq \rho_0^{ll'} \leq \rho_2^{ll'}$ for all $l$ and $l'$, meaning our model specification assumes the between-period ICCs are less than or equal to the within-period ICCs, and is a generalization of the common assumption in the analysis of SW-CRTs with a single outcome. Finally, in situations where limited information is available for ICC determination, an investigator could set the between-period ICCs to be a certain ratio of the within-period ICCs, referred to as the cluster autocorrelation coefficient (CAC).\citep{hooper2016sample,martin2016intra} Of note, when the variance components of the cluster-by-time random interactions are zero, $\bSigma_s=\boldsymbol{0}$, the MLMM \eqref{eq:MLMM} can be considered as a direct extension of the LMM developed by Hussey and Hughes \citep{Hussey2007} when there is more than one outcome.

\begin{table}[htbp]
    \caption{\label{t:ICCs}Definition of intracluster correlation coefficients (ICCs) with total variance for the $l$-th outcome denoted by $\sigma^2_{yl}=\sigma^2_{bl}+\sigma^2_{sl}+\sigma^2_{\epsilon l}$.}
    \centering
    {\fontsize{10.5}{11.5}\selectfont
    \begin{tabular}{lll}
        \toprule
        ICC & Definition & Expression\\
        \midrule
        $\rho_0^l$ & within-period endpoint-specific ICC & $\text{corr}(Y_{ijkl},Y_{ijk'l})=(\sigma^2_{bl}+\sigma^2_{sl})/\sigma^2_{yl}$\\
        $\rho_1^l$ & between-period endpoint-specific ICC & $\text{corr}(Y_{ijkl},Y_{ij'k'l})=\sigma^2_{bl}/\sigma^2_{yl}$\\
        $\rho_0^{ll'}$ & within-period between-endpoint ICC & $\text{corr}(Y_{ijkl},Y_{ijk'l'})=(\sigma_{bll'}+\sigma_{sll'})/(\sigma_{yl}\sigma_{yl'})$\\
        $\rho_1^{ll'}$ & between-period between-endpoint ICC & $\text{corr}(Y_{ijkl},Y_{ij'k'l'})=\sigma_{bll'}/(\sigma_{yl}\sigma_{yl'})$\\
        $\rho_2^{ll'}$ & intra-subject ICC & $\text{corr}(Y_{ijkl},Y_{ijkl'})=(\sigma_{bll'}+\sigma_{sll'}+\sigma_{\epsilon ll'})/(\sigma_{yl}\sigma_{yl'})$\\
        \bottomrule
    \end{tabular}
    }
\end{table}

Before we detail the design considerations for a cross-sectional SW-CRT with multivariate outcomes, we briefly describe the fitting strategies of the MLMM \eqref{eq:MLMM}. We adopt the expectation-maximization (EM) algorithm where random effects are treated as missing data. The EM algorithm is an iterative approach that includes two steps: (1) generating the expected values of the random effects given the current parameter estimates and (2) using those expected values to generate updated parameter estimates using score functions. The first step fittingly refers to the expectation stage and the second step to the maximization stage since the score function for each parameter produces a value that maximizes the likelihood. The EM algorithm iterates between these two steps until convergence is met, usually defined as a negligible change in the likelihood (i.e. $10^{-5}$). Let $\btheta=(\bbeta^\top,\bsigma^\top)^\top$, where $\bbeta$ is the vector of all fixed effects and $\bsigma$ is the vector of all variance components (unique components that make up $\bSigma_b$, $\bSigma_s$, and $\bSigma_\epsilon$), denote our set of parameters we wish to estimate. Also let $\D_{ij}$ denote the design matrix for the fixed effects, $\bbeta$, for cluster $i$ at period $j$. We can express the fully observed likelihood of our MLMM using 
\begin{align*}
     f(\Y,\bb,\s|\btheta)&=\prod_{i=1}^I\prod_{j=1}^T\prod_{k=1}^{N_{ij}} f(\Y_{ijk}|\bb_{i},\s_{ij};\btheta)f(\bb_{i}|\btheta)f(\s_{ij}|\btheta),
\end{align*}
where $f(\Y_{ijk}|\bb_{i},\s_{ij};\btheta)$, $f(\bb_{i}|\btheta)$ and $f(\s_{ij}|\btheta)$ are the conditional multivariate normal density of the outcome, multivariate normal density for the random cluster effects and multivariate normal density for the random cluster-by-time interactions (detailed expressions are provided in Web Appendix A). To generate the score functions for the maximization step we take the partial derivative of our log-likelihood with respect to a particular parameter, set the expression equal to zero, and solve for that parameter giving us the following (derivations provided in Web Appendix A)
\begin{align*}
     S(\bbeta)&=\left(\sum_{i=1}^I\sum_{j=1}^T\sum_{k=1}^{N_{ij}}\D_{ij}^\top\bSigma_{\epsilon}^{-1}\D_{ij}\right)^{-1}\sum_{i=1}^I\sum_{j=1}^T\sum_{k=1}^{N_{ij}}\D_{ij}^\top\bSigma_{\epsilon}^{-1}(\Y_{ijk}-\bb_i-\s_{ij})\\
     S(\bSigma_b)&=\dfrac{1}{I}\sum_{i=1}^I\bb_{i}\bb_{i}^\top,~~~~~
     S(\bSigma_s)=\dfrac{1}{IT}\sum_{i=1}^I\sum_{j=1}^T\s_{ij}\s_{ij}^\top\\
     S(\bSigma_{\epsilon})&=\dfrac{1}{\sum_{i=1}^I\sum_{j=1}^TN_{ij}}\sum_{i=1}^I\sum_{j=1}^T\sum_{k=1}^{N_{ij}}(\Y_{ijk}-\D_{ij}\bbeta-\bb_i-\s_{ij})(\Y_{ijk}-\D_{ij}\bbeta-\bb_i-\s_{ij})^\top.
\end{align*}
Next we need to generate the expected value of the random effects and crossproducts for the expectation step. To achieve this, we first re-parameterize our MLMM using the equivalent expression, $\Y_{ijk}=\D_{ij}^\top\bbeta+\M_{ij}^\top\bphi_i+\bepsilon_{ijk}$, where $\bphi_i=(\bb_i^\top,\s_{i1}^\top,\ldots,\s_{iT}^\top)^\top$ is the vector of random effects for cluster $i$ and follows a multivariate normal distribution with mean $\0_{L(T+1)\times1}$ and covariance $\bSigma_\phi=\text{diag}\left(\bSigma_b,\I_T \otimes \bSigma_s\right)$ where $\I_T$ denotes a $T\times T$ identity matrix. Further, $\M_{ij}$ is the design matrix of the random effects, $\bphi_i$. Under this equivalent parameterization, the posterior distribution of the random effects for cluster $i$ is $f(\bphi_i|\Y_{i};\btheta) = {f(\Y_{i},\bphi_i|\btheta)}/{f(\Y_{i}|\btheta)}$, which can be shown to be a multivariate normal density, based on which the expected sufficient statistics can be constructed based on 
\begin{align*}
     \Exp(\bphi_i|\Y_i,\btheta)&=\left(\bSigma_\phi^{-1}+N_{ij}\sum_{j=1}^T\M_{ij}\bSigma_\epsilon^{-1}\M_{ij}^\top\right)^{-1}\sum_{j=1}^T\sum_{k=1}^{N_{ij}}\M_{ij}\bSigma_{\epsilon}^{-1}(\Y_{ijk}-\D_{ij}^\top\bbeta)\\
     \text{V}(\bphi_i|\Y_i,\btheta)&=\left(\bSigma_\phi^{-1}+N_{ij}\sum_{j=1}^T\M_{ij}\bSigma_\epsilon^{-1}\M_{ij}^\top\right)^{-1}\\
     \Exp(\bphi_i\bphi_i^\top|\Y_i,\btheta)&=\text{V}(\bphi_i|\Y_i,\btheta) + \Exp(\bphi_i|\Y_i,\btheta)\Exp(\bphi_i|\Y_i,\btheta)^\top,
 \end{align*}
where $\Y_i=(\Y_{i11}^\top,\ldots,\Y_{iTN_{iT}}^\top)^\top$ is a vector of all outcomes measured in cluster $i$. In Web Appendix A, we summarize the operational details for the EM algorithm. Once the algorithm converges, the standard errors for model parameter estimators are obtained from numerically differentiating the log-likelihood function evaluated at the model parameter estimates. In Section \ref{s:simulation}, we demonstrate via simulations that the EM approach can provide nominal type I error rate and comparable empirical power relative to the formula predictions, in the context of SW-CRTs. For ease of reference, the development of our sample size methodology relies on matrix notation which is described in Table \ref{t:notation}.
\begin{table}
    \caption{\label{t:notation}Glossary of notation.}
    \centering
    {\fontsize{9}{10}\selectfont
    \begin{tabular}{ll}
        \toprule
        Notation & Meaning\\
        \midrule
        $L$,$I$,$T$,$N$ & Number of outcomes, clusters, periods, subjects per cluster (balanced design)\\
        $\Sigma_b$,$\Sigma_s$,$\Sigma_\epsilon$,$\Sigma_\gamma$ & cluster, cluster-period, error, within-subject (closed-cohort) variance components\\
        $\I_u$,$\J_u$ & $u\times u$ identity matrix, $u\times u$ matrix of ones\\
        $A \otimes B$ & Kronecker product of the matrices $A$ and $B$\\
        $A \circ B$ & Hadamard (element-wise) product of the matrices $A$ and $B$\\
        $U$,$V$,$W$ & $\sum_{i=1}^I\sum_{j=1}^TX_{ij}$, $\sum_{i=1}^I\left(\sum_{j=1}^TX_{ij}\right)^2$, $\sum_{j=1}^T\left(\sum_{i=1}^IX_{ij}\right)^2$\\
        $\bLambda_y$ & $\text{diag}(\sigma_{y1}^2,\ldots,\sigma_{yL}^2)$\\
        $\bsigma_{\epsilon}$ & $(\sigma_{\epsilon 1},\ldots,\sigma_{\epsilon L})^\top$\\
        $\bomega$ & $\left(\sigma_{y1}(1-\rho_0^1)^{1/2},\ldots, \sigma_{yL}(1-\rho_0^L)^{1/2}\right)^\top$ or $\left(\sigma_{y1}\left(1-\rho_0^1+\rho_1^1-\rho_2^1\right)^{1/2},\ldots, \sigma_{yL}\left(1-\rho_0^L+\rho_1^L-\rho_2^L\right)^{1/2}\right)^\top$ (closed-cohort)\\
        $\lambda_1$ & $1-\rho_0$ or $1-\rho_0+\rho_1-\rho_2$ (closed-cohort)\\
        $\lambda_2$ & $1+(N-1)\rho_0-N\rho_1$\\
        $\lambda_3$ & $1+(N-1)\rho_0+(T-1)N\rho_1$ or $1+(N-1)(\rho_0-\rho_1)-\rho_2$ (closed-cohort)\\
        $\lambda_4$ & $1+(N-1)\rho_0+(T-1)(N-1)\rho_1+(T-1)\rho_2$\\
        $\tau_2$ & $(N-1)\rho_{00}-N\rho_{11}+\rho_2$\\
        $\tau_3$ & $(N-1)\rho_{00}-N\rho_{11}+\rho_2+TN\rho_{11}$ or $\rho_{2,0}-\rho_{2,1}+(N-1)(\rho_{00}-\rho_{11})$ (closed-cohort)\\
        $\tau_4$ & $\rho_{2,0}-\rho_{2,1}+(N-1)(\rho_{00}-\rho_{11})+T\left(\rho_{2,1}+(N-1)\rho_{11}\right)$\\
        $\bGamma_0$,$\bGamma_1$ & $\begin{pmatrix}
        \rho_0^{1} & \rho_0^{12} & \dots & \rho_0^{1L}\\
        \rho_0^{12} & \rho_0^{2} & \dots & \rho_0^{2L}\\
        \vdots & \vdots & \ddots & \vdots\\
        \rho_0^{1L} & \rho_0^{2L} & \dots & \rho_0^{L}\\
        \end{pmatrix},~~
        \begin{pmatrix}
        \rho_1^{1} & \rho_1^{12} & \dots & \rho_1^{1L}\\
        \rho_1^{12} & \rho_1^{2} & \dots & \rho_1^{2L}\\
        \vdots & \vdots & \ddots & \vdots\\
        \rho_1^{1L} & \rho_1^{2L} & \dots & \rho_1^{L}\\
        \end{pmatrix}$\\
        $\bGamma_2$ & $\begin{pmatrix}
        1 & \rho_2^{12} & \dots & \rho_2^{1L}\\
        \rho_2^{12} & 1 & \dots & \rho_2^{2L}\\
        \vdots & \vdots & \ddots & \vdots\\
        \rho_2^{1L} & \rho_2^{2L} & \dots & 1\\
        \end{pmatrix} or ~~
        \begin{pmatrix}
        1 & \rho_{2,0}^{12} & \dots & \rho_{2,0}^{1L}\\
        \rho_{2,0}^{12} & 1 & \dots & \rho_{2,0}^{2L}\\
        \vdots & \vdots & \ddots & \vdots\\
        \rho_{2,0}^{1L} & \rho_{2,0}^{2L} & \dots & 1\\
        \end{pmatrix}$ (closed-cohort)\\
        $\bGamma_{2'}$ & $\begin{pmatrix}
        \rho_2^1 & \rho_{2,1}^{12} & \dots & \rho_{2,1}^{1L}\\
        \rho_{2,1}^{12} & \rho_2^2 & \dots & \rho_{2,1}^{2L}\\
        \vdots & \vdots & \ddots & \vdots\\
        \rho_{2,1}^{1L} & \rho_{2,1}^{2L} & \dots & \rho_2^L\\
        \end{pmatrix}$\\
        \bottomrule
    \end{tabular}
    }
\end{table}


\section{Design Considerations: Power Calculation with Multivariate Outcomes}
\label{s:power}

\subsection{Variance of the Intervention Effect Estimator and Power Formula}\label{sec:general}
For testing the intervention effect on outcome $l$, we consider the Wald $t$-statistic for $\delta_l$, defined as $w_l=\hat{\delta}_l/\hat{\sigma}_{\delta_l}$ where $\hat{\sigma}_{\delta_l}$ denotes the estimated standard error of the intervention effect estimator from the MLMM \eqref{eq:MLMM}. In the design stage, we can express $\sigma_{\delta_l}$ using the Feasible Generalized Least Square (FGLS) formula, which is given by $\left(\sum_{i=1}^I\Z_i^\top\widetilde{\V}_i^{-1}\Z_i\right)^{-1}$, where $\Z_i$ is the design matrix for the fixed effects at the cluster-level and $\widetilde{\V}_i$ is the covariance matrix for the $L$ cluster-period mean outcomes. A cluster-period means approach is applicable since the fixed effects of model \eqref{eq:MLMM} only depend on each cluster-period and has been shown to be equivalent to an individual-level approach.\citep{li2021marginal,davis2021sample,ligeneralizing2022} For simplicity, we assume a balanced design where each cluster recruits the same number of subjects in each period ($N_{ij}=N$). In Web Appendix B, we generate the inverse of $\widetilde{\V}_i$ as
\begin{align*}
    \widetilde{\V}_i^{-1}&=\I_T \otimes \left(\bSigma_s + \dfrac{1}{N}\bSigma_\epsilon\right)^{-1} + \J_T \otimes \dfrac{1}{T}\left[\left(T\bSigma_b+\bSigma_s+\dfrac{1}{N}\bSigma_\epsilon\right)^{-1}-\left(\bSigma_s+\dfrac{1}{N}\bSigma_\epsilon\right)^{-1}\right],
\end{align*}
where 
$\J_u$ denotes a $u \times u$ matrix of ones. If we let $\Z_i=(\I_T,\X_{i}) \otimes \I_L$ where $\X_i$ denotes the randomization schedule for cluster $i$, then the bottom-right $L \times L$ matrix of the FGLS estimator will be the covariance matrix for the $L$ intervention effect estimators. Using the FGLS estimator and our expressions for $\widetilde{\V}_i^{-1}$ and $\Z_i$, we derive in Web Appendix B an expression for the $L \times L$ covariance matrix of the intervention effect estimators as
\begin{align}\label{eq:vard}
    \resizebox{0.77\textwidth}{!}{$\bOmega_\delta=IT\Bigg[\left(ITU-TW+U^2-IV\right)\left(\bSigma_s+\dfrac{1}{N}\bSigma_\epsilon\right)^{-1}
    -\left(U^2-IV\right)\left(T\bSigma_b+\bSigma_s+\dfrac{1}{N}\bSigma_\epsilon\right)^{-1}\Bigg]^{-1},$}
\end{align}
where $U=\sum_{i=1}^I\sum_{j=1}^TX_{ij}$, $V=\sum_{i=1}^I\left(\sum_{j=1}^TX_{ij}\right)^2$, and $W=\sum_{j=1}^T\left(\sum_{i=1}^IX_{ij}\right)^2$ are typical design constants that only depend on the randomization sequence of intervention indicators. From Table \ref{t:ICCs}, we can map the variance component parameters to the set of unique ICC parameters by observing $\sigma_{bl}^2=\sigma_{yl}^2\rho_1^l$, $\sigma_{bll^{\prime}}=\sigma_{yl}\sigma_{yl^\prime}\rho_1^{ll^{\prime}}$, $\sigma_{sl}^2=\sigma_{yl}^2(\rho_0^l-\rho_1^l)$, $\sigma_{sll^{\prime}}=\sigma_{yl}\sigma_{yl^\prime}(\rho_0^{ll^{\prime}}-\rho_1^{ll^{\prime}})$, $\sigma_{\epsilon l}^2=\sigma_{yl}^2(1-\rho_0^l)$ and $\sigma_{\epsilon ll^{\prime}}=\sigma_{yl}\sigma_{yl^\prime}(\rho_2^{ll^{\prime}}-\rho_0^{ll^{\prime}})$. Defining the diagonal matrix of outcome variances as $\bLambda_y=\text{diag}(\sigma_{y1}^2,\ldots,\sigma_{yL}^2)$, we can further rewrite the covariance matrix of the intervention effect estimators as
\begin{align}\label{eq:vard_icc}
    \bOmega_\delta=&\frac{IT}{N}\bLambda_y^{1/2}\Bigg[\left(ITU-TW+U^2-IV\right)\left(\bGamma_2-N\bGamma_1+(N-1)\bGamma_0\right)^{-1}\nonumber\\
        &-\left(U^2-IV\right)\left(\bGamma_2+(T-1)N\bGamma_1+(N-1)\bGamma_0\right)^{-1}\Bigg]^{-1}\bLambda_y^{1/2},
\end{align}
where $\bGamma_0$, $\bGamma_1$, $\bGamma_2$ are the within-period ICC matrix, between-period ICC matrix and intra-subject ICC matrix across $L$ endpoints, defined as
$$\bGamma_0=\begin{pmatrix}
\rho_0^{1} & \rho_0^{12} & \dots & \rho_0^{1L}\\
\rho_0^{12} & \rho_0^{2} & \dots & \rho_0^{2L}\\
\vdots & \vdots & \ddots & \vdots\\
\rho_0^{1L} & \rho_0^{2L} & \dots & \rho_0^{L}\\
\end{pmatrix},~~
\bGamma_1=\begin{pmatrix}
\rho_1^{1} & \rho_1^{12} & \dots & \rho_1^{1L}\\
\rho_1^{12} & \rho_1^{2} & \dots & \rho_1^{2L}\\
\vdots & \vdots & \ddots & \vdots\\
\rho_1^{1L} & \rho_1^{2L} & \dots & \rho_1^{L}\\
\end{pmatrix},~~
\bGamma_2=\begin{pmatrix}
1 & \rho_2^{12} & \dots & \rho_2^{1L}\\
\rho_2^{12} & 1 & \dots & \rho_2^{2L}\\
\vdots & \vdots & \ddots & \vdots\\
\rho_2^{1L} & \rho_2^{2L} & \dots & 1\\
\end{pmatrix}.$$

We denote diagonal elements of $\bOmega_\delta$ as $\sigma^2_{\delta_l}=\text{var}(\hat{\delta_l})$ and off-diagonal elements as $\sigma_{\delta_{ll'}}$. Under the case of a univariate outcome, such that $\bSigma_b$, $\bSigma_s$, and $\bSigma_\epsilon$ are scalars, we show in Web Appendix B that our covariance expressions \eqref{eq:vard} and \eqref{eq:vard_icc} reduce to the variance under the Hooper and Girling model for cross-sectional SW-CRTs.\citep{hooper2016sample,Girling2016} 
We further explore the relationship between CAC and $\lim_{N\to\infty}\text{var}(\hat{\delta_l})$ and find that the importance of differentiating the within-period and between-period ICCs in SW-CRTs with univariate outcomes still holds under multivariate outcomes. 
In particular, we found that
\begin{align*}
    \lim_{N\rightarrow\infty}\bOmega_\delta\propto\bLambda_y^{1/2}\Bigg[\left(ITU-TW+U^2-IV\right)\left(\bGamma_0-\bGamma_1\right)^{-1}
        -\left(U^2-IV\right)\left(\bGamma_0+(T-1)\bGamma_1\right)^{-1}\Bigg]^{-1}\bLambda_y^{1/2},
\end{align*}
which is not equal to a zero matrix unless $\bGamma_0=\bGamma_1$. Thus, allowing some differentiation ($\text{CAC}<1$) leads to a non-zero limiting variance and reduces the possibility of an underpowered trial.

Based on the closed-form variance expression $\bOmega_\delta$, we focus on power analyses when the $L$ outcomes are co-primary outcomes, in which case the test rejects the null when the intervention leads to meaningful changes on all $L$ outcomes. Alternatively, when there is interest in testing whether at least one outcome is affected by the intervention (an omnibus test) or testing whether the treatment effect is homogeneous for all $L$ outcomes (testing for treatment effect homogeneity), one could combine $\bOmega_\delta$ with the generic power formulas in Yang et al.\citep{yang2022power} to derive a suitable power formula. To proceed with the co-primary outcome example, we notice that the joint distribution of the Wald $t$-statistics, $\W=(w_1,\ldots,w_L)^\top=(\hat{\delta}_1/\hat{\sigma}_{\delta_1},\ldots, \hat{\delta}_L/\hat{\sigma}_{\delta_L})^\top$, asymptotically follows a multivariate normal distribution with mean $\bbbeta=(\delta_1/\sigma_{\delta_1},\ldots,\delta_L/\sigma_{\delta_L})^\top$ and covariance matrix $\bOmega_{\W}$ which is characterized by ones on the diagonal and $\sigma^2_{\delta_{ll'}}/(\sigma_{\delta_{l}}\sigma_{\delta_{l'}})$ on the off-diagonal; we will refer to this joint distribution using $f_{\W}(\W)$. In practice, investigators could alternatively assume $f_{\W}(\W)$ follows a multivariate $t$-distribution with degrees of freedom characterized by the total number of clusters ($I$) and co-primary endpoints ($L$), specifically $I-2L$, to account for the uncertainty in estimating the covariance components and better control the type I error, especially if the study has a limited number of clusters.\citep{li2020power}

\begin{figure}
\centering
\includegraphics[width=\textwidth]{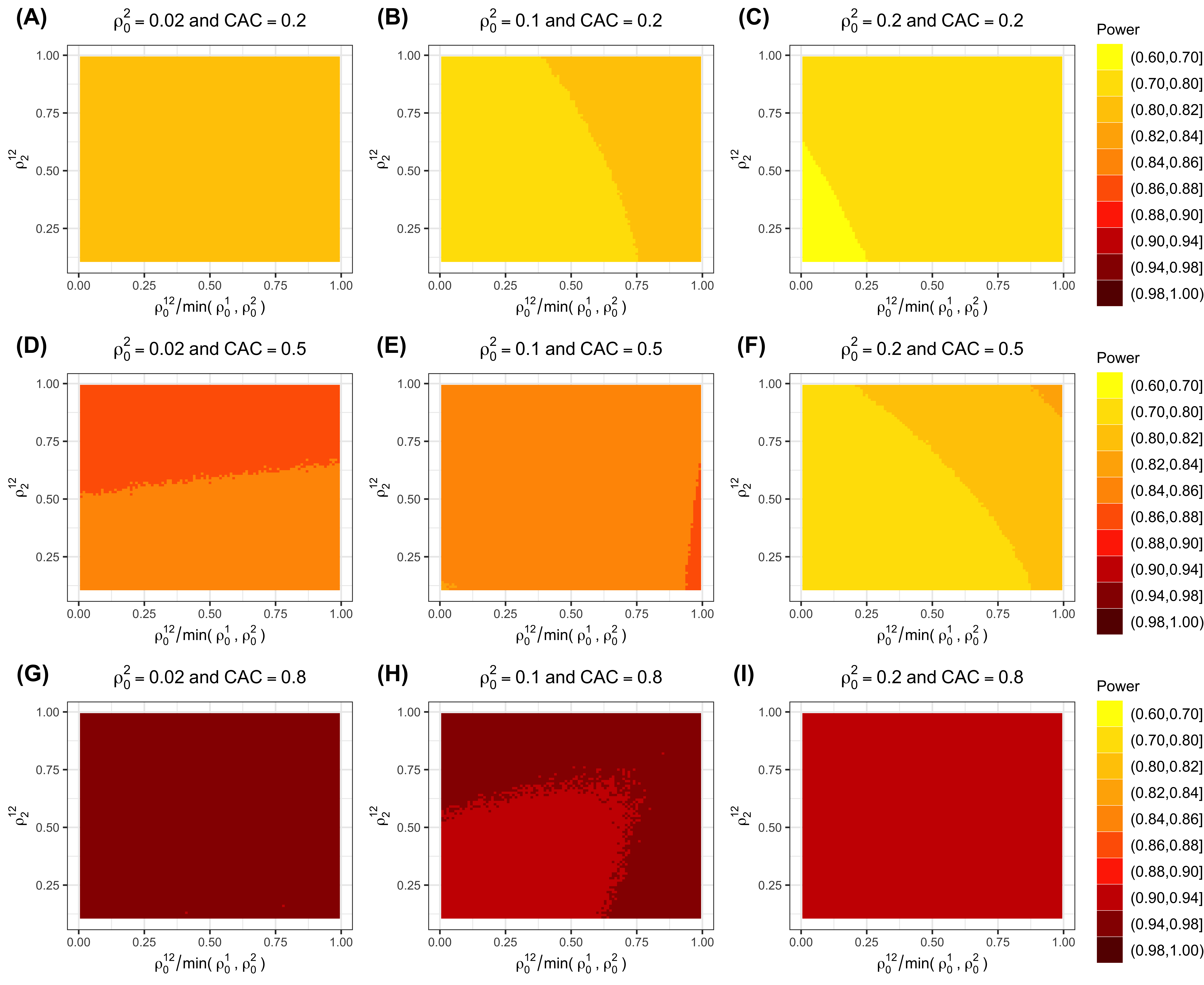}
\caption{\label{f:contour}Relationship between ICCs and power with $L=2$ co-primary endpoints. ICCs include the within-period endpoint-specific ICCs ($\rho_0^1=0.1$, $\rho_0^2$), between-period endpoint-specific ICCs ($\rho_1^1$, $\rho_1^2$), within-period and between-period between-endpoint ICCs ($\rho_0^{12}$, $\rho_1^{12}$), and intra-subject ICC ($\rho_2^{12}$). Various $\rho_2^{12}$ values are shown on the y-axis and various $\rho_0^{12}$ specifications are shown on the x-axis as a ratio of $\text{min}(\rho_0^1,\rho_0^2)$. Between-period ICCs are defined by CAC. Darker colors denote higher power.}
\end{figure}

We can now explicitly define our null and alternative hypotheses and generate power predictions. In this study, we focus on the scenario where all $L$ outcomes are of primary interest and generate power using the IU-test given by
\begin{align*}
    H_0\colon\; \exists\; l \in \{1,\ldots,L\}\; \text{s.t}\; \delta_l=0,~~~~~~~~~
    H_1\colon\; \forall\; l \in \{1,\ldots,L\}\; \delta_l > 0,
\end{align*}
which has been frequently utilized in the context of co-primary endpoints to avoid inflated type I error.\citep{chuang2007challenge,sozu2010sample} Under the IU-test we can generate our power estimate using
\begin{align}\label{eq:power}
    \text{power}&=\text{Prob}\left(\mathcal{R}=\bigcap_{l=1}^L\{w_l>c_l\}\right)=\int_{c_1}^\infty \ldots \int_{c_L}^\infty f_{\W}(w_1,\ldots,w_L)d_1 \ldots d_L,
\end{align}
where $\mathcal{R}$ denotes the rejection region, $\bc=\{c_1,\ldots,c_L\}$ is the set of critical values for rejection, and $f_{\W}(\W)$ follows a multivariate $t$-distribution. A common and simple approach for specifying $\bc$ is to use $c_1=\ldots=c_L=t_\alpha(I-2L)$ where $t_\alpha(I-2L)$ is the $(1-\alpha)$th quantile of the univariate $t$-distribution since this approach leads to a type I error rate strictly below $\alpha$ over the composite null space.\citep{sozu2010sample,sozu2012sample,li2020power} Only under the extreme case of one endpoint showing no intervention effect and all remaining endpoints showing large intervention effects, will the maximum type I error rate be achieved.\citep{berger1982multiparameter} The relationship between power and each of the ICCs is shown in Figure \ref{f:contour}. As has been demonstrated for SW-CRTs with one primary outcome, higher values of the within-period endpoint-specific ICC ($\rho_0^l$) and typically lower values of the between-period endpoint-specific ICC ($\rho_1^l$) yield lower power values. Further, Figure \ref{f:contour} suggests that lower values of between-endpoint ICCs both within and between-period ($\rho_0^{12}$, $\rho_1^{12}$) as well as the intra-subject ICC ($\rho_2^{12}$) lead to lower power estimates. Although we include this exploration of the relationship between ICCs and power, we caution that this relationship could be more complex. For instance, under a SW-CRT design with a single outcome, Davis-Plourde et al.\cite{davis2021sample} found that power had a monotone relationship with the within-period ICCs and a non-monotone relationship with the between-period ICCs. Finally, although we focus on determining power, the sample size corresponding to a target level of power can be determined by solving for $I$ or $N$ using any standard iterative algorithm. 

Our alternative hypothesis under the IU-test assumes that the desired effect of the intervention is an increase in all measured outcomes. If the desired intervention effect is a decrease in all measured outcomes, then our IU-test can still be used by changing the definition of each outcome. The power formula \eqref{eq:power} can also be used for a mixture of superiority and noninferiority tests across the $L$ outcomes through the specification of the critical values, $\bc$. For example, for a noninferiority test on the $l$-th endpoint, i.e. $H_{0l}\colon \delta_{l} < \lambda_l$ versus $H_{1l}\colon \delta_l \geq \lambda_l$ with $\lambda_l$ denoting the noninferiority margin, the test statistic becomes $w_l=(\hat{\delta_l}-\lambda_l)/\hat{\sigma}_{\delta_l}$. 

\subsection{Common Intervention Effects}
In some cases the global impact of intervention instead of individual effects may be of interest, and simplification of the above power formula may be of interest to facilitate study designs. Common intervention effects are frequently assumed in the social sciences, mental health, and meta-analysis.\citep{hedges2014statistical} A usual assumption under the common intervention effects model is that the standardized intervention effects are the same.\citep{turner2006modelling} If such an assumption is valid, then incorporating common effects in the model can lead to more precise estimates. In this case, we can modify model \eqref{eq:MLMM} to include a common intervention effect for all $L$ outcomes using a standardized scale
\begin{align}\label{eq:MLMM_cte}
    \Y_{ijk}&=(\bbeta'_{0}+\bbeta'_{j}+X_{ij}\delta'+\bb'_{i}+\s'_{ij})\circ\bsigma_{\epsilon}+\bepsilon_{ijk},
\end{align}
where $\bsigma_{\epsilon}=(\sigma_{\epsilon 1},\ldots,\sigma_{\epsilon L})^\top$, $\bb'_{i}$ follows a multivariate normal distribution denoted by $\mathcal{N}(\0_{L \times 1}, \bSigma'_b)$, and $\s'_{ij}$ follows a multivariate normal distribution denoted by $\mathcal{N}(\0_{L \times 1}, \bSigma'_s)$. Under this specification, an overall intervention effect is given by $\delta'$ which corresponds to individual intervention effects on the original outcome scales using the relationship $\delta_l=\sigma_{\epsilon l}\delta'$. Similarly, remaining model effects can be translated to the original scale using $\beta_{0l}=\sigma_{\epsilon l}\beta'_{0l}$, $\beta_{jl}=\sigma_{\epsilon l}\beta'_{jl}$, $b_{il}=\sigma_{\epsilon l}b'_{il}$, and $s_{ijl}=\sigma_{\epsilon l}s'_{ijl}$. An alternative approach to model \eqref{eq:MLMM_cte} is to standardize relative to the total variance ($\sigma_{yl}^2$), however, this approach may be less attractive in SW-CRTs where the variance components of the random effects can be challenging to estimate with a limited number of clusters.\citep{turner2006modelling} Under model \eqref{eq:MLMM_cte}, we show in Web Appendix C that the variance of the common intervention effect estimator is
\begin{align}\label{eq:vard_icc_ct}
    \text{var}(\hat{\delta'})=&\frac{IT}{N}\Bigg[\left(ITU-TW+U^2-IV\right)\bomega^\top\bLambda_y^{-1/2}\left(\bGamma_2-N\bGamma_1+(N-1)\bGamma_0\right)^{-1}\bLambda_y^{-1/2}\bomega\nonumber\\
        &-\left(U^2-IV\right)\bomega^\top\bLambda_y^{-1/2}\left(\bGamma_2+(T-1)N\bGamma_1+(N-1)\bGamma_0\right)^{-1}\bLambda_y^{-1/2}\bomega\Bigg]^{-1},
\end{align}
where $\bomega=\left(\sigma_{y1}(1-\rho_0^1)^{1/2},\ldots, \sigma_{yL}(1-\rho_0^L)^{1/2}\right)^\top$ and can be used in standard power procedures for SW-CRTs with a single outcome. For example, if we are interested in a one-sided Wald test such that $H_0\colon \delta'=0$ versus $H_1\colon \delta' > 0$, then we can use \eqref{eq:vard_icc_ct} along with  $\mbox{power} \approx 1-\Phi_{t}\left(t_{\alpha}(\text{DF});\text{DF},|\delta'|/\sqrt{\mbox{var}(\hat{\delta'})}\right)$,
where $\Phi_{t}(t;\text{DF},\Lambda)$ is the cumulative $t$-distribution function with DF degrees of freedom and noncentrality parameter $\Lambda$, and $t_{\alpha}(\text{DF})$ is the $(1-\alpha)$th quantile of the central $t$-distribution. We can specify the degrees of freedom as a function of the total number of clusters ($I$) minus $L$ period effects and one intervention effect ($\text{DF}=I-L-1$), as an extension of the degrees of freedom proposed in Ford et al.\citep{ford2020maintaining} and Li \citep{Li2019decay} with a univariate outcome.

\subsection{Common ICC Values across Endpoints} 
In many SW-CRTs, it may not be unreasonable to assume a common ICC across endpoints, since cluster and other characteristics inducing similarity within clusters can plausibly affect multiple outcomes measured in those clusters.\citep{turner2006modelling} Futhermore, endpoint-specific ICC values may not be known precisely at the time of sample size calculation and may instead, rely on rules of thumb based on the type of outcome. If the ICC is assumed to be common across the $L$ multivariate endpoints, then there are no longer outcome-specific ICCs and only five ICCs regardless of $l$, such that $\rho_0^l=\rho_0\text{ and } \rho_1^l=\rho_1\: \forall l$ and $\rho_0^{ll'}=\rho_{00}, \rho_1^{ll'}=\rho_{11},\text{ and } \rho_2^{ll'}=\rho_2\: \forall l,l'$. 
Based on the mapping between the ICC parameters and variance components in Table \ref{t:ICCs}, the common ICC assumption leads to simplification of the variance expression $\bOmega_\delta$ by defining the three key ICC matrices with their explicit simple exchangeable forms: $\bGamma_0=(\rho_0-\rho_{00})\I_L+\rho_{00}\J_L$, $\bGamma_1=(\rho_1-\rho_{11})\I_L+\rho_{11}\J_L$ and $\bGamma_2=(1-\rho_{2})\I_L+\rho_{2}\J_L$. 
Plugging in these explicit forms into variance expression \eqref{eq:vard_icc}, we show in Web Appendix D that we can express the covariance matrix of the intervention effect estimators as
\begin{align}\label{eq:vard_icc2}
    \bOmega_\delta=&\frac{IT}{N}\bLambda_y^{1/2}\Bigg[\left(ITU-TW+U^2-IV\right)\left((\lambda_2-\tau_2)\I_L+\tau_2\J_L\right)^{-1}
        -\left(U^2-IV\right)\left((\lambda_3-\tau_3)\I_L+\tau_3\J_L\right)^{-1}\Bigg]^{-1}\bLambda_y^{1/2},
\end{align}
where $\lambda_2=1+(N-1)\rho_0-N\rho_1$ and $\lambda_3=1+(N-1)\rho_0+(T-1)N\rho_1$ are two distinct eigenvalues of the (endpoint-specific) nested exchangeable correlation structure \citep{LiTurnerPreisser2018} defined for cross-sectional SW-CRTs with a univariate outcome, and $\tau_2=(N-1)\rho_{00}-N\rho_{11}+\rho_2$, $\tau_3=\tau_2+TN\rho_{11}$ characterize the impact of the three between-endpoint ICCs on the variance of intervention effect estimators through the MLMM. In the special case where all endpoints are completely independent such that $\rho_{00}=\rho_{11}=\rho_2=0$, $\bOmega_\delta$ becomes a diagonal matrix and each element becomes identical to the variance expression developed in Hooper et al.\citep{hooper2016sample} and Girling et al.\citep{Girling2016} for cross-sectional SW-CRTs with a univariate outcome. However, we can obtain additional insights into the general form of the diagonal element of $\bOmega_\delta$ and summarize the results in the following Theorem (proof in Web Appendix D).
\begin{theorem}\label{thm1}
Under the parsimonious parameterization with common ICC values across endpoints, the $l$-th diagonal element of $\bOmega_\delta$ can be further written in the following analytical form
\begin{align*}
\text{var}(\hat{\delta}_l)=&\frac{(IT/N)\sigma_{yl}^2}{\left(ITU-TW+U^2-IV\right)(\lambda_3-\tau_3)-\left(U^2-IV\right)(\lambda_2-\tau_2)}\times \\
&\hspace{-2cm}\frac{\left(ITU-TW+U^2-IV\right)\lambda_2(\lambda_3-\tau_3)\left\{\lambda_3+(L-1)\tau_3\right\}-\left(U^2-IV\right)\lambda_3(\lambda_2-\tau_2)\left\{\lambda_2+(L-1)\tau_2\right\}}{\left(ITU-TW+U^2-IV\right)\left\{\lambda_3+(L-1)\tau_3\right\}-\left(U^2-IV\right)\left\{\lambda_2+(L-1)\tau_2\right\}}.
\end{align*}
Furthermore, denote the variance of the $l$-th intervention effect estimator based on a univariate Hooper and Girling model \citep{hooper2016sample,Girling2016} is 
\begin{align*}
\text{var}^{\text{HG}}(\hat{\delta}_l)=\frac{(IT/N)\sigma_{yl}^2\lambda_2\lambda_3}{\left(ITU-TW+U^2-IV\right)\lambda_3-\left(U^2-IV\right)\lambda_2},
\end{align*}
and $\text{var}(\hat{\delta}_l)\leq \text{var}^{\text{HG}}(\hat{\delta}_l)$ for any set of valid design parameters, with equality holds when $\tau_2\lambda_3=\tau_3\lambda_2$ or $\rho_{00}=\rho_{11}=\rho_2=0$ (a special case when $\tau_2\lambda_3=\tau_3\lambda_2$).
\end{theorem}
Theorem \ref{thm1} shows that the diagonal element of $\bOmega_\delta$ is always smaller than the existing variance expression developed in Hooper et al.\citep{hooper2016sample} and Girling et al.\citep{Girling2016} for compatible set of design parameters, explicitly revealing that modeling multivariate outcomes through MLMM will frequently lead to improved efficiency for estimating the endpoint-specific treatment effect, compared to separate LMM analyses. This is in sharp contrast to the previous results developed for designing parallel-arm cluster randomized trials, where MLMM and separate LMM analyses lead to the same asymptotic efficiency for estimating the endpoint-specific treatment effect when the cluster sizes are equal \citep{yang2022power}. Therefore, in a stepped wedge design, modeling multivariate outcomes through MLMM will frequently lead to a reduced sample size and larger power for testing the endpoint-specific treatment effect. Finally, the advantages of assuming common ICCs increases with $L$, in that the number of ICCs being estimated is significantly reduced as $L\rightarrow \infty$ and ICC estimation will likely be more precise. When the ICC parameters are anticipated to differ across endpoints, the more general expression developed in Section \ref{sec:general} should be considered.

\subsection{Common Intervention Effects and Common ICC Values across Endpoints}
Lastly, it may be of interest to assume both common ICCs and common intervention effects. For a cross-sectional design, we can use model \eqref{eq:MLMM_cte} and further simplify the variance of the intervention effect estimator \eqref{eq:vard_icc_ct} by replacing the ICC matrices with their exchangeable forms giving us (derivation in Web Appendix E)
\begin{align}\label{eq:vard_both_icc_cte}
    \text{var}(\hat{\delta'})&=\dfrac{(IT/(LN\lambda_1))(\lambda_2+(L-1)\tau_2)(\lambda_3+(L-1)\tau_3)}{(ITU-TW+U^2-IV)(\lambda_3+(L-1)\tau_3)-(U^2-IV)(\lambda_2+(L-1)\tau_2)},
\end{align}
where $\lambda_2$, $\lambda_3$, and $\tau_3$ are the same as defined previously and $\lambda_1=1-\rho_0$ is an additional distinct eigenvalue of the (endpoint-specific) nested exchangeable correlation structure \citep{LiTurnerPreisser2018} defined for cross-sectional SW-CRTs with a univariate outcome. This variance expression can be used in standard power procedures for SW-CRTs with a single outcome. If a common intervention effect and common ICCs are expected, then the advantage of using the common intervention effect model \eqref{eq:MLMM_cte} should be an increase in the precision of the intervention effect estimate. In the following Theorem, we formally compare the variance formula for both common ICCs and a common intervention effect \eqref{eq:vard_both_icc_cte} to the variance formula assuming common ICCs (Theorem \ref{thm1}) and summarize our results (proof in Web Appendix E).
\begin{theorem}\label{thm2}
Under the parsimonious parameterization with common ICC values and a common intervention effect across endpoints, the variance of the $l$-th intervention effect estimator (unscaled) under model \eqref{eq:MLMM_cte}, i.e. $\delta_l=\sigma_{yl}\lambda_1^{1/2}\delta'$, is
\begin{align*}
    \text{var}^{\text{both}}(\hat{\delta}_l)&=\dfrac{(IT/(LN))\sigma_{yl}^2(\lambda_2+(L-1)\tau_2)(\lambda_3+(L-1)\tau_3)}{(ITU-TW+U^2-IV)(\lambda_3+(L-1)\tau_3)-(U^2-IV)(\lambda_2+(L-1)\tau_2)}.
\end{align*}
As shown in Theorem \ref{thm1}, under the parsimonious parameterization with common ICC values across endpoints, the $l$-th diagonal element of $\bOmega_\delta$ is denoted by
\begin{align*}
\text{var}^{\text{ICC}}(\hat{\delta}_l)=&\frac{(IT/N)\sigma_{yl}^2}{\left(ITU-TW+U^2-IV\right)(\lambda_3-\tau_3)-\left(U^2-IV\right)(\lambda_2-\tau_2)}\times \\
&\hspace{-2cm}\frac{\left(ITU-TW+U^2-IV\right)\lambda_2(\lambda_3-\tau_3)\left\{\lambda_3+(L-1)\tau_3\right\}-\left(U^2-IV\right)\lambda_3(\lambda_2-\tau_2)\left\{\lambda_2+(L-1)\tau_2\right\}}{\left(ITU-TW+U^2-IV\right)\left\{\lambda_3+(L-1)\tau_3\right\}-\left(U^2-IV\right)\left\{\lambda_2+(L-1)\tau_2\right\}},
\end{align*}
and $\text{var}^{\text{both}}(\hat{\delta}_l) < \text{var}^{\text{ICC}}(\hat{\delta}_l)$ for any set of valid design parameters.
\end{theorem}
In Theorem \ref{thm1} we showed that assuming common ICCs often leads to more precise estimates of the intervention effect estimator and in Theorem \ref{thm2} we showed that additionally assuming common intervention effects further improves that precision due to information borrowing across different endpoints to estimate a single treatment effect. 

\subsection{Extensions to Closed-Cohort Designs}
Thus far we have focused on cross-sectional SW-CRTs with continuous multivariate outcomes. If individuals within a cluster are followed over time, corresponding to a closed-cohort design, then additional intra-subject ICCs are necessary to take into account the correlation of repeated measurements within the same subject.\citep{LiTurnerPreisser2018} In terms of our model \eqref{eq:MLMM}, this means including additional random effects to account for repeated measures
\begin{align}\label{eq:MLMM_cc}
    \Y_{ijk}&=\bbeta_{0}+\bbeta_{j}+\bdelta X_{ij}+\bb_{i}+\s_{ij}+\bgamma_{ik}+\bepsilon_{ijk},
\end{align}
where $\bgamma_{ik}=(\gamma_{ik1},\ldots,\gamma_{ikL})^\top$ is a vector of random subject-level effects and remaining parameters are the same as described previously. We assume $\bgamma_{ik}$ follows a multivariate normal distribution denoted by $\mathcal{N}(\0_{L \times 1}, \bSigma_{\gamma})$. We denote the diagonal elements of $\bSigma_{\gamma}$ as $\sigma^2_{\gamma l}$ and off-diagonal elements as $\sigma_{\gamma ll'}$ giving us a total of $L(L+1)/2$ variance components for specifying $\bSigma_{\gamma}$. Similar to model \eqref{eq:MLMM}, we assume $\bb_{i}$, $\s_{ij}$, $\bgamma_{ik}$, and $\bepsilon_{ijk}$ are independent and place no further restrictions on $\bSigma_b$, $\bSigma_s$, $\bSigma_{\gamma}$, and $\bSigma_{\epsilon}$ except to be positive definite. Under this specification, the marginal variance for the $l$-th outcome is $\sigma^2_{yl}=\sigma^2_{bl}+\sigma^2_{sl}+\sigma^2_{\gamma l}+\sigma^2_{\epsilon l}$ and the first four ICC definitions under model \eqref{eq:MLMM} remain the same (only the marginal variance of the outcome changes). Under model \eqref{eq:MLMM_cc}, the fifth ICC and two additional intra-subject ICCs defined under model \eqref{eq:MLMM} becomes, 
\begin{enumerate}
    \item[(5)] $\rho_{2,0}^{ll'}=\text{corr}(Y_{ijkl},Y_{ijkl'})=(\sigma_{bll'}+\sigma_{sll'}+\sigma_{\gamma ll'}+\sigma_{\epsilon ll'})/(\sigma_{yl}\sigma_{yl'})$ denoting the intra-subject within-period between-endpoint ICC or for brevity, the within-period intra-subject ICC;
    \item[(6)] $\rho_{2,1}^{ll'}=\text{corr}(Y_{ijkl},Y_{ij'kl'})=(\sigma_{bll'}+\sigma_{\gamma ll'})/(\sigma_{yl}\sigma_{yl'})$ denoting the intra-subject between-period between-endpoint ICC or for brevity, the between-period intra-subject ICC;
    \item[(7)] $\rho_2^l=\text{corr}(Y_{ijkl},Y_{ij'kl})=(\sigma^2_{bl}+\sigma^2_{\gamma l})/\sigma^2_{yl}$ denoting the intra-subject between-period endpoint-specific ICC or for brevity, the intra-subject endpoint-specific ICC.
\end{enumerate}
\begin{table}[htbp]
    \caption{\label{t:ICCs_cc}Definition of intracluster correlation coefficients (ICCs) under a closed-cohort design with total variance for the $l$-th outcome denoted by $\sigma^2_{yl}=\sigma^2_{bl}+\sigma^2_{sl}+\sigma^2_{\gamma l}+\sigma^2_{\epsilon l}$.}
    \centering
    \begin{tabular}{lll}
        \hline
        ICC & Definition & Expression\\
        \hline
        $\rho_0^l$ & within-period endpoint-specific ICC & $\text{corr}(Y_{ijkl},Y_{ijk'l})=(\sigma^2_{bl}+\sigma^2_{sl})/\sigma^2_{yl}$\\
        $\rho_1^l$ & between-period endpoint-specific ICC & $\text{corr}(Y_{ijkl},Y_{ij'k'l})=\sigma^2_{bl}/\sigma^2_{yl}$\\
        $\rho_0^{ll'}$ & within-period between-endpoint ICC & $\text{corr}(Y_{ijkl},Y_{ijk'l'})=(\sigma_{bll'}+\sigma_{sll'})/(\sigma_{yl}\sigma_{yl'})$\\
        $\rho_1^{ll'}$ & between-period between-endpoint ICC & $\text{corr}(Y_{ijkl},Y_{ij'k'l'})=\sigma_{bll'}/(\sigma_{yl}\sigma_{yl'})$\\
        $\rho_{2,0}^{ll'}$ & within-period intra-subject ICC & $\text{corr}(Y_{ijkl},Y_{ijkl'})=(\sigma_{bll'}+\sigma_{sll'}+\sigma_{\gamma ll'}+\sigma_{\epsilon ll'})/(\sigma_{yl}\sigma_{yl'})$\\
        $\rho_{2,1}^{ll'}$ & between-period intra-subject ICC & $\text{corr}(Y_{ijkl},Y_{ij'kl'})=(\sigma_{bll'}+\sigma_{\gamma ll'})/(\sigma_{yl}\sigma_{yl'})$\\
        $\rho_2^l$ & intra-subject endpoint-specific ICC & $\text{corr}(Y_{ijkl},Y_{ij'kl})=(\sigma^2_{bl}+\sigma^2_{\gamma l})/\sigma^2_{yl}$\\
        \hline
    \end{tabular}
\end{table}
Table \ref{t:ICCs_cc} provides a summary of these ICC parameters under a closed-cohort design. The same properties of symmetry and degeneracy under model \eqref{eq:MLMM} also applies to all ICCs under model \eqref{eq:MLMM_cc}. Specifically, under model \eqref{eq:MLMM_cc} we have symmetry in that $\rho_0^{ll'}=\rho_0^{l'l}$, $\rho_1^{ll'}=\rho_1^{l'l}$, $\rho_{2,0}^{ll'}=\rho_{2,0}^{l'l}$, and $\rho_{2,1}^{ll'}=\rho_{2,1}^{l'l}$, and degeneracy such that $\rho_0^{ll}=\rho_0^l$, $\rho_1^{ll}=\rho_1^l$, $\rho_{2,0}^{ll}=1$, and $\rho_{2,1}^{ll}=\rho_2^l$. Our ICC definitions also implicitly assume $\rho_1^l \leq \rho_0^l$, $\rho_1^{ll'} \leq \rho_0^{ll'} \leq \rho_{2,0}^{ll'}$, and $\rho_{2,1}^{ll'} \leq \rho_{2,0}^{ll'}$ for all $l$ and $l'$, meaning our model specification again assumes the between-period ICCs are less than or equal to the within-period ICCs. Furthermore, we show in Web Appendix F that the covariance matrix of the intervention effects under model \eqref{eq:MLMM_cc} is
\begin{align}\label{eq:vard_cc}
    \bOmega_\delta &=\dfrac{IT}{N}\bLambda_y^{1/2}\Bigg[(ITU-TW+U^2-IV)\left\{(N-1)(\bGamma_0-\bGamma_1)+\bGamma_2-\bGamma_{2'}\right\}^{-1}\nonumber\\
    &-(U^2-IV)\left\{(T-1)(N-1)\bGamma_1+(T-1)\bGamma_{2'}+(N-1)\bGamma_0+\bGamma_2\right\}^{-1}\Bigg]^{-1}\bLambda_y^{1/2},
\end{align}
where $\bGamma_{2}$ is slightly modified to reflect the change in the fifth ICC and $\bGamma_{2'}$ is an additional intra-subject ICC matrix that takes into account the closed-cohort design defined by
$$\bGamma_2=\begin{pmatrix}
1 & \rho_{2,0}^{12} & \dots & \rho_{2,0}^{1L}\\
\rho_{2,0}^{12} & 1 & \dots & \rho_{2,0}^{2L}\\
\vdots & \vdots & \ddots & \vdots\\
\rho_{2,0}^{1L} & \rho_{2,0}^{2L} & \dots & 1\\
\end{pmatrix},~~
\bGamma_{2'}=\begin{pmatrix}
\rho_2^1 & \rho_{2,1}^{12} & \dots & \rho_{2,1}^{1L}\\
\rho_{2,1}^{12} & \rho_2^2 & \dots & \rho_{2,1}^{2L}\\
\vdots & \vdots & \ddots & \vdots\\
\rho_{2,1}^{1L} & \rho_{2,1}^{2L} & \dots & \rho_2^L\\
\end{pmatrix}.$$
This covariance matrix expression, $\bOmega_\delta$, can be used in conjunction with \eqref{eq:power} to estimate power and sample size for designing closed-cohort SW-CRTs with multivariate endpoints.

\subsubsection{Common Intervention Effects under Closed-Cohort Designs}
We can extend our model for estimating a common intervention effect to a closed-cohort design by incorporating our additional random effect in model \eqref{eq:MLMM_cte} giving us
\begin{align}\label{eq:MLMM_cte_cc}
    \Y_{ijk}&=(\bbeta'_{0}+\bbeta'_{j}+X_{ij}\delta'+\bb'_{i}+\s'_{ij}+\bgamma'_{ik})\circ\bsigma_{\epsilon}+\bepsilon_{ijk},
\end{align}
where $\bgamma'_{ik}$ follows a multivariate normal distribution denoted by $\mathcal{N}(\0_{L \times 1}, \bSigma'_\gamma)$ and remaining effects are the same as previously described. Effects can be translated to the original scale using the same properties described earlier in addition to $\gamma_{ijl}=\sigma_{\epsilon l}\gamma'_{ijl}$. Under model \eqref{eq:MLMM_cte_cc}, we show in Web Appendix G that the variance of the common intervention effect estimator is
\begin{align}\label{eq:vard_ct_cc}
    \text{var}(\hat{\delta'})=&\frac{IT}{N}\Bigg[\left(ITU-TW+U^2-IV\right)\bomega^\top\bLambda_y^{-1/2}\left\{(N-1)(\bGamma_0-\bGamma_1)+\bGamma_2-\bGamma_{2'}\right\}^{-1}\bLambda_y^{-1/2}\bomega\nonumber\\
        &-\left(U^2-IV\right)\bomega^\top\bLambda_y^{-1/2}\left\{(T-1)(N-1)\bGamma_1+(T-1)\bGamma_{2'}+(N-1)\bGamma_0+\bGamma_2\right\}^{-1}\bLambda_y^{-1/2}\bomega\Bigg]^{-1},
\end{align}
where $\bomega=\left(\sigma_{y1}\left(1-\rho_0^1+\rho_1^1-\rho_2^1\right)^{1/2},\ldots, \sigma_{yL}\left(1-\rho_0^L+\rho_1^L-\rho_2^L\right)^{1/2}\right)^\top$ and can be used in standard power procedures for SW-CRTs with a single outcome.

\subsubsection{Common ICC Values across Endpoints under Closed-Cohort Designs}
The common ICC assumption again leads to simplification of the variance expression $\bOmega_\delta$ by defining the three key ICC matrices with their explicit simple exchangeable forms:
\begin{align*}
\bGamma_0=&(\rho_0-\rho_{00})\I_L+\rho_{00}\J_L\\
\bGamma_1=&(\rho_1-\rho_{11})\I_L+\rho_{11}\J_L\\
\bGamma_2=&(1-\rho_{2,0})\I_L+\rho_{2,0}\J_L\\
\bGamma_2'=&(\rho_2-\rho_{2,1})\I_L + \rho_{2,1}\J_L.
\end{align*}
Plugging in these explicit forms into our current variance expression \eqref{eq:vard_cc} gives us (derivation in Web Appendix H)
\begin{align}\label{eq:vard_icc_cc}
    \bOmega_\delta &=\dfrac{IT}{N}\bLambda_y^{1/2}\Bigg[(ITU-TW+U^2-IV)\big\{(\lambda_3-\tau_3)\I_L+\tau_3\J_L\big\}^{-1}-(U^2-IV)\big\{(\lambda_4-\tau_4)\I_L+\tau_4\J_L \big\}^{-1}\Bigg]^{-1}\bLambda_y^{1/2},
\end{align}
where $\lambda_3=1+(N-1)(\rho_0-\rho_1)-\rho_2$ and $\lambda_4=1+(N-1)\rho_0+(T-1)(N-1)\rho_1+(T-1)\rho_2$ are two distinct eigenvalues of the (endpoint-specific) block exchangeable correlation structure.\citep{hooper2016sample,Girling2016,li2021mixed} Further, $\tau_3=\rho_{2,0}-\rho_{2,1}+(N-1)(\rho_{00}-\rho_{11})$ and $\tau_4=\tau_3+T\left(\rho_{2,1}+(N-1)\rho_{11}\right)$ characterize the impact of the between-endpoint ICCs on the variance of intervention effect estimators through the MLMM. In the special case where all endpoints are completely independent such that $\rho_{00}=\rho_{11}=\rho_{2,0}=\rho_{2,1}=0$, $\bOmega_\delta$ becomes a diagonal matrix and each element becomes identical to the variance expression developed in Hooper et al.\citep{hooper2016sample}, Girling and Hemming\citep{Girling2016}, and Li et al.\citep{li2021mixed} for closed-cohort SW-CRTs with a univariate outcome. Concentrating on the diagonal elements of $\bOmega_\delta$ gives us the following Theorem as an extension of Theorem \ref{thm1} (proof in Web Appendix H).
\begin{theorem}\label{thm3}
Under the parsimonious parameterization with common ICC values across endpoints and a closed-cohort design, the $l$-th diagonal element of $\bOmega_\delta$ can be further written in the following analytical form
\begin{align*}
\text{var}(\hat{\delta}_l)=&\frac{(IT/N)\sigma_{yl}^2}{\left(ITU-TW+U^2-IV\right)(\lambda_4-\tau_4)-\left(U^2-IV\right)(\lambda_3-\tau_3)}\times \\
&\hspace{-2cm}\frac{\left(ITU-TW+U^2-IV\right)\lambda_3(\lambda_4-\tau_4)\left\{\lambda_4+(L-1)\tau_4\right\}-\left(U^2-IV\right)\lambda_4(\lambda_3-\tau_3)\left\{\lambda_3+(L-1)\tau_3\right\}}{\left(ITU-TW+U^2-IV\right)\left\{\lambda_4+(L-1)\tau_4\right\}-\left(U^2-IV\right)\left\{\lambda_3+(L-1)\tau_3\right\}}.
\end{align*}
Furthermore, denote the variance of the $l$-th intervention effect estimator based on a univariate Hooper and Girling model \citep{hooper2016sample,Girling2016,li2021mixed} is 
\begin{align*}
\text{var}^{\text{HG}}(\hat{\delta}_l)=\frac{(IT/N)\sigma_{yl}^2\lambda_3\lambda_4}{\left(ITU-TW+U^2-IV\right)\lambda_4-\left(U^2-IV\right)\lambda_3}
\end{align*}
and $\text{var}(\hat{\delta}_l)\leq \text{var}^{\text{HG}}(\hat{\delta}_l)$ for any set of valid design parameters, with equality holds when $\tau_3\lambda_4=\tau_4\lambda_3$ or $\rho_{00}=\rho_{11}=\rho_{2,0}=\rho_{2,1}=0$ (a special case when $\tau_3\lambda_4=\tau_4\lambda_3$).
\end{theorem}
Similar to Theorem \ref{thm1}, Theorem \ref{thm3} shows that the diagonal element of $\bOmega_\delta$ is always smaller than the existing variance expression developed in Hooper et al.\cite{hooper2016sample}, Girling and Hemming\cite{Girling2016}, and Li et al.\cite{LiTurnerPreisser2018} for compatible set of design parameters. Thus, the improved efficiency under SW-CRTs with a cross-sectional design remains under a closed-cohort design.

\subsubsection{Common Intervention Effects and Common ICC Values across Endpoints under Closed-Cohort Designs}
Under a closed-cohort design, we can use model \eqref{eq:MLMM_cte_cc} and further simplify our variance expression \eqref{eq:vard_ct_cc} using the ICC matrices exchangeable forms giving us (derivation in Web Appendix I)
\begin{align*}
    \text{var}(\hat{\delta'})&=\dfrac{(IT/(LN\lambda_1))(\lambda_3+(L-1)\tau_3)(\lambda_4+(L-1)\tau_4)}{(ITU-TW+U^2-IV)(\lambda_4+(L-1)\tau_4)-(U^2-IV)(\lambda_3+(L-1)\tau_3)},
\end{align*}
where $\lambda_3$, $\lambda_4$, $\tau_3$, and $\tau_4$ are the same as previously defined and $\lambda_1=1-\rho_0+\rho_1-\rho_2$ is a distinct eigenvalue of the (endpoint-specific) block exchangeable correlation structure \citep{LiTurnerPreisser2018} defined for closed-cohort SW-CRTs with a univariate outcome. This variance expression can be used in standard power procedures for SW-CRTs with a single outcome. In the following Theorem, we extend Theorem \ref{thm2} under a cross-sectional design to a closed-cohort design (proof in Web Appendix I).
\begin{theorem}\label{thm4}
Under the parsimonious parameterization with common ICC values and a common intervention effect across endpoints under a closed-cohort design, the variance of the $l$-th intervention effect estimator (unscaled) under model \eqref{eq:MLMM_cte_cc}, i.e. $\delta_l=\sigma_{yl}\lambda_1^{1/2}\delta'$, is
\begin{align*}
    \text{var}^{\text{both}}(\hat{\delta}_l)&=\dfrac{(IT/(LN))\sigma_{yl}^2(\lambda_3+(L-1)\tau_3)(\lambda_4+(L-1)\tau_4)}{(ITU-TW+U^2-IV)(\lambda_4+(L-1)\tau_4)-(U^2-IV)(\lambda_3+(L-1)\tau_3)}.
\end{align*}
As shown in Theorem \ref{thm3}, under the parsimonious parameterization with common ICC values across endpoints, the $l$-th diagonal element of $\bOmega_\delta$ is denoted by
\begin{align*}
\text{var}^{\text{ICC}}(\hat{\delta}_l)=&\frac{(IT/N)\sigma_{yl}^2}{\left(ITU-TW+U^2-IV\right)(\lambda_4-\tau_4)-\left(U^2-IV\right)(\lambda_3-\tau_3)}\times \\
&\hspace{-2cm}\frac{\left(ITU-TW+U^2-IV\right)\lambda_3(\lambda_4-\tau_4)\left\{\lambda_4+(L-1)\tau_4\right\}-\left(U^2-IV\right)\lambda_4(\lambda_3-\tau_3)\left\{\lambda_3+(L-1)\tau_3\right\}}{\left(ITU-TW+U^2-IV\right)\left\{\lambda_4+(L-1)\tau_4\right\}-\left(U^2-IV\right)\left\{\lambda_3+(L-1)\tau_3\right\}},
\end{align*}
and $\text{var}^{\text{both}}(\hat{\delta}_l) < \text{var}^{\text{ICC}}(\hat{\delta}_l)$ for any set of valid design parameters.
\end{theorem}
Just as we saw under cross-sectional designs, assuming common ICCs often leads to more precise estimates of the intervention effect estimator (Theorem \ref{thm3}) and additionally assuming common intervention effects further improves that precision (Theorem \ref{thm4}) under closed-cohort designs. 


\section{Application to Data Example}
\label{s:application}

\subsection{Shared Decision-Making in Interprofessional Home Care Teams (IP-SDM) Trial}\label{s:IP-SDM}
The shared decision-making in interprofessional home care teams (IP-SDM) study is a SW-CRT focused on evaluating the implementation of shared decision-making in interprofessional home care teams caring for elderly clients and their caregivers in Quebec, Canada.\citep{adisso2022shared} Nine Health and Social Services Centers (HSSCs) were randomized to one of four possible sequences ($T=5$) with two HSSCs per sequence with the exception of sequence two which had three HSSCs. At each time period eight different elderly clients ($N=8$) were sampled from each HSSC. The primary endpoint of the IP-SDM trial was the binary decision to stay at home or to move with scores on various questionnaires as secondary outcomes. A secondary outcome of interest was health-related quality of life  of the elderly clients measured using the Nottingham Health Profile (NHP). The investigators chose two relevant and equally important subscales of the NHP, namely social isolation and emotional reactions. Each subscale includes multiple yes/no questions which are then weighted and summed to generate a total score ranging from $0$ to $100$.

\subsection{Sample Size and Power Considerations in the Context of IP-SDM Study}\label{s:IP-SDM_app}
To illustrate our power and sample size methodology, we use data from the IP-SDM study to inform the design of a future cross-sectional SW-CRT to study the effect of the shared decision making model on social isolation and emotional reactions as two co-primary endpoints. The investigators plan to use the same number of periods $(T=5)$, but need to determine the number of clusters $(I)$ and cluster-period size $(N)$ required to achieve at least 80\% power at the 5\% nominal level under the IU-test as described in Section \ref{sec:general}. We also assume an equal number of clusters per sequence, i.e. $I$ will be a multiple of 4. First, we use data from the IP-SDM study to obtain plausible estimates of the ICCs for the future study. To estimate the ICCs, we fit model \eqref{eq:MLMM} using the EM algorithm with social isolation designated as the first outcome and emotional reactions as the second outcome and use the relationship between the estimated variance components and ICCs (Table \ref{t:ICCs}). Using the results of the model, we estimated $\sigma_{b1}^2=0.01$, $\sigma_{s1}^2=3.71$, and $\sigma_{\epsilon 1}^2=607.41$ giving us a total variance of $\sigma^2_{y1}=611.13$ for social isolation. For emotional reactions, we estimated $\sigma_{b2}^2=4.74$, $\sigma_{s2}^2=15.69$, and $\sigma_{\epsilon 2}^2=675.30$ giving us a total variance of $\sigma^2_{y2}=695.73$. For the between-outcome variance components, we estimated $\sigma_{b12}=-0.05$, $\sigma_{s12}=-1.23$, and $\sigma_{\epsilon 12}=377.27$. The negative estimates of $\sigma_{b12}$ and $\sigma_{s12}$ imply negative ICC values which may be reasonable in certain situations, but given the relatively small sample size of the IP-SDM trial, were considered as sampling error for our purposes. Given that a positive correlation is plausible  between the the NHP subscores, we instead set $\sigma_{b12}=0$ and $\sigma_{s12}=0$ corresponding to a between-endpoint ICC (both within-period and between-period) of zero. In other words, we assume that there is no correlation between different individuals within the same cluster measured on different outcomes. Now that we have all of our variance components, we can use Table \ref{t:ICCs} to generate the ICC values. Specifically, we estimate the endpoint-specific ICCs for social isolation to be $\rho_0^1=0.006$ (within-period) and $\rho_1^1=0.00002$ (between-period). For emotional reactions, we estimate the endpoint-specific ICCs to be $\rho_0^2=0.029$ (within-period) and $\rho_1^2=0.0068$ (between-period). We further estimate the between-endpoint ICCs to be $\rho_0^{12}=\rho_1^{12}=0$ (within-period and between-period) and the intra-subject ICC to be $\rho_2^{12}=0.58$. For demonstration, we assume that a clinically relevant effect size of intervention is $\delta_1=0.30 \times \sigma_{y1}$ and $\delta_2=0.35 \times \sigma_{y2}$, i.e. that the shared decision making intervention leads to an increase in each of the quality of life subscales. Using our ICCs, design parameters, and equations \eqref{eq:vard_icc} and \eqref{eq:power}, we estimate that $I=16$ HSSCs (clusters) with $N=12$ clients per HSSC per period are needed to achieve $86.3\%$ power based on the IU-test. 

\subsection{Sensitivity Analysis to ICC Assumptions}
We assess the sensitivity of our power estimate to ICC specification in Table \ref{t:app_sens}. For simplification, we specify all between-period ICCs using CAC and consider varying levels (0.0, 0.2, 0.5, and 0.8). Further, since our point estimates for $\sigma_{b12}$ and $\sigma_{s12}$ were negative, implying the between-outcome ICCs ($\rho_0^{12}$ and $\rho_1^{12}$) are negative, we include such scenarios in our sensitivity analysis in addition to scenarios of positive correlation. For all remaining ICCs, we explored values that were 20\%, 40\%, and 60\% lower and higher than the current specification. Overall, we found that our predicted power (86.3\%) was robust to the ICC values considered in our sensitivity analysis. Specifically, the predicted power was slightly higher for lower values of within-period endpoint-specific ICCs ($\rho_0^1$, $\rho_0^2$) and for higher values of the intra-subject ICC ($\rho_2^{12}$). Further, the predicted power was fairly constant for various between-period ICCs (denoted by CAC) and for the within-period between-endpoint ICC ($\rho_0^{12}$).
\begin{table}
    \caption{\label{t:app_sens}Sensitivity of ICC specification in our application to the IP-SDM study. ICCs include the within-period endpoint-specific ICCs ($\rho_0^1=0.006$, $\rho_0^2=0.029$), the within-period between-endpoint ICC ($\rho_0^{12}=0$), and the intra-subject ICC ($\rho_2^{12}=0.58$). Between-period ICCs are defined by CAC. Assuming a total of $I=16$ HSSCs (clusters) with $N=12$ clients per HSSC per period produced a predicted power of 86.3\%.}
    \centering
    \begin{tabular}{cccccc}
        \toprule
        $\rho_2^{12}$ & $\rho_0^1$ & $\rho_0^2$ & $\rho_0^{12}$ & CAC & power (\%)\\
        \midrule
        0.58 & 0.006 & 0.029 & 0 & {0.0} & 86.9\\
        0.58 & 0.006 & 0.029 & 0 & {0.2} & 86.2\\
        0.58 & 0.006 & 0.029 & 0 & {0.5} & 86.0\\
        0.58 & 0.006 & 0.029 & 0 & {0.8} & 86.5\\
        0.58 & 0.006 & 0.029 & {-0.004} & 0.2 & 86.1\\
        0.58 & 0.006 & 0.029 & {-0.002} & 0.2 & 86.1\\
        0.58 & 0.006 & 0.029 & {0.002} & 0.2 & 86.2\\
        0.58 & 0.006 & 0.029 & {0.004} & 0.2 & 86.3\\
        0.58 & 0.006 & {0.012} & 0 & 0.2 & 88.6\\
        0.58 & 0.006 & {0.017} & 0 & 0.2 & 87.7\\
        0.58 & 0.006 & {0.023} & 0 & 0.2 & 87.0\\
        0.58 & 0.006 & {0.035} & 0 & 0.2 & 85.3\\
        0.58 & 0.006 & {0.041} & 0 & 0.2 & 84.4\\
        0.58 & 0.006 & {0.046} & 0 & 0.2 & 83.7\\
        0.58 & {0.002} & 0.029 & 0 & 0.2 & 87.2\\
        0.58 & {0.004} & 0.029 & 0 & 0.2 & 86.6\\
        0.58 & {0.005} & 0.029 & 0 & 0.2 & 86.3\\
        0.58 & {0.007} & 0.029 & 0 & 0.2 & 85.9\\
        0.58 & {0.008} & 0.029 & 0 & 0.2 & 85.7\\
        0.58 & {0.010} & 0.029 & 0 & 0.2 & 85.5\\
        {0.23} & 0.006 & 0.029 & 0 & 0.2 & 85.1\\
        {0.35} & 0.006 & 0.029 & 0 & 0.2 & 85.3\\
        {0.46} & 0.006 & 0.029 & 0 & 0.2 & 85.7\\
        {0.70} & 0.006 & 0.029 & 0 & 0.2 & 86.7\\
        {0.81} & 0.006 & 0.029 & 0 & 0.2 & 87.3\\
        {0.93} & 0.006 & 0.029 & 0 & 0.2 & 88.4\\
        \bottomrule
    \end{tabular}
\end{table}

\subsection{Additional Comparative Analyses}
For comparison purposes, if an investigator is interested in testing whether either endpoint is significantly different from zero (an omnibus test), then with the same number of HSSCs ($I=16$) and clients per HSSC per period ($N=12$) researchers have around the same level of power, 86.5\%, to detect much smaller effect sizes of $\delta_1=0.052 \times \sigma_{y1}$ and $\delta_2=0.102 \times \sigma_{y2}$. This is expected as the omnibus test would be rejecting more frequently than the IU-test due to a larger space of alternative hypotheses. 

In addition, we carried out a comparison between the stepped wedge design and a parallel-arm design, assuming equal total sample sizes. Specifically, if we use the same design parameter inputs for the IP-SDM study (standardized effect size of (0.30, 0.35)) but assume a typical parallel-arm cluster randomized trial design (with the same total number of subjects but without requiring the information on between-period ICCs), then the power would be 91.5\% compared to 86.3\% under a stepped wedge design using the IU-test. Since the IU-test rejects the null based on the endpoint-specific test statistic, its operating characteristics are expected to be more similar to the conventional Wald-test for a single endpoint. From this point of view, the above comparison result is in agreement with the findings in Hemming and Taljaard\citep{hemming2016sample} that a parallel-arm design can be slightly more efficient than a stepped wedge design (with a single endpoint) when the ICC is small. However, the omnibus test is based on a quadratic test statistic involving the covariance matrix of the treatment effect estimator and thus may be more affected by the magnitude of the between-period ICCs under a stepped wedge design. Indeed, with the omnibus test, if we use the same standardized effect sizes, (0.052, 0.102), the power under a parallel-arm design is only 12.0\% compared to 86.5\% under a stepped wedge design. This finding confirms the different behaviour between the IU-test and omnibus test that was previously identified under a parallel-arm design.\citep{yang2022power} We acknowledge that this is only an empirical comparison under the IP-SDM example, and a formal comparison between parallel-arm design and stepped wedge design with multiple outcomes is yet to be investigated in future work. Finally, when evaluating alternative designs for a specific study, the decision to adopt a stepped wedge design is often not exclusively based on power and can include other practical or administrative considerations; see, for example, broad justifications for stepped wedge designs detailed in Hemming and Taljaard.\cite{hemming2020reflection}


\section{A Simulation Study}
\label{s:simulation}

For further illustration, we validate our proposed methodology for estimating power under the IU-test using simulations. We consider two ($L=2$) normally distributed multivariate outcomes under a cross-sectional design. For each endpoint-specific ICC, we chose within-period ICCs within commonly reported ranges for parallel-arm CRTs, $\rho_0^l=\{0.02,0.1,0.2\}$, and between-period ICCs using a CAC of 0.5, $\rho_1^l=\{0.01,0.05,0.1\}$. For between-endpoint ICCs, we set the within-period ICC using $\rho_0^{ll'}=0.5\times\text{min}(\rho_0^l,\rho_0^{l'})=\{0.01,0.05,0.1\}$, the between-period ICC using a CAC of 0.5, $\rho_1^{ll'}=\{0.005,0.025,0.05\}$, and we considered small to large values of the intra-subject ICC, $\rho_2^{ll'}=\{0.2,0.5,0.8\}$. We considered all possible combinations of our five ICC parameters giving a total of 27 scenarios. To generate continuous multivariate outcomes we restrict the total variance of each endpoint ($\sigma^2_{y1}=\sigma^2_{y2}=4$) which allows us to compute individual variance components based on the ICC values (Table \ref{t:ICCs}), and use the MLMM \eqref{eq:MLMM} introduced in Section \ref{s:model}, $\Y_{ijk}=\bbeta_{0}+\bbeta_{j}+X_{ij}\bdelta+\bb_{i}+\s_{ij}+\bepsilon_{ijk}$. As shown in equation \eqref{eq:vard}, the covariance matrix of the intervention effects is time invariant, therefore we only assume a minor and common increasing secular trend for each endpoint with $\beta_{01}=\beta_{02}=0$ and $\beta_{(j+1)1}-\beta_{j1}=\beta_{(j+1)2}-\beta_{j2}=0.1\times(0.5)^{j-1}$ for $j\geq 1$. Our simulations also assumed a standard SW-CRT design such that an equal number of clusters are randomly assigned to each sequence. Motivated by the findings in a systematic review of SW-CRTS \citep{Grayling2017b}, we varied the number of clusters ($I$) between 8 and 30, the number of periods ($T$) between 3 and 5, and set an upper limit of 25 for the cluster-period size ($N$). We considered standardized intervention effect sizes of $\delta_l/\sigma_{yl} \in [0.1,1]$. Exact parameter values were chosen to ensure at least 80\% power based on a one-sided nominal 5\% level Wald test. To compute the predicted power we used \eqref{eq:vard_icc} and \eqref{eq:power} with critical values $c_1=c_2=t_{\alpha}(I-4)$. The empirical power of the Wald test was determined by the proportion correctly rejecting $H_0$, $\Ind\left\{\bigcap_{l=1}^2\{w_l>c_l\}\right\}=1$, over 1000 simulated SW-CRTs, when the MLMM parameters are estimated by the EM algorithm. Agreement between the empirical and predicted power was used to assess the accuracy of our proposed method. Finally, since the maximum error rate is supposed to be achieved only when one of the true treatment effects is zero, we assessed the empirical type I error rate by setting $\delta_1/\sigma_{y1}=0$ only and then setting $\delta_2/\sigma_{y2}=0$ only to confirm the validity of the Wald test. In Web Table 1, we also provide the empirical type I error when treatment effects on both endpoints are set to zero.

In Table \ref{t:sim_results} we present the empirical power and type I error rate and predicted power of the Wald test for each scenario. Overall, the empirical type I error rate was conservative, and differences between empirical and predicted power were small, ranging between -1.8\% and 3.9\%. Thus, our power and sample size methodology either closely matched or slightly underestimated the true power (therefore at worst conservative). Therefore, our proposed method based on asymptotics is accurate and can be used to design cross-sectional SW-CRTs with multivariate continuous outcomes without resorting to computationally exhaustive simulation-based calculations. 
\begin{table}
    \caption{\label{t:sim_results}Estimated required number of clusters $I$, subjects per cluster-period $N$, periods $T$, empirical type I error when only setting the first effect to zero and when only setting the second effect to zero $(e_1,e_2)$, empirical power $\bar{\psi}$, and predicted power $\psi$ obtained from power formula for given effect size $\delta^l/\sigma_{\delta l}$, within-period and between-period endpoint-specific ICCs ($\rho_0^l,\rho_1^l$), within-period and between-period between-endpoint ICCs ($\rho_0^{ll'},\rho_1^{ll'}$), and intra-subject ICC ($\rho_2^{ll'}$) assuming a CAC of 0.5 with $L=2$ co-primary endpoints.}
    \centering
    {\fontsize{10.5}{11.5}\selectfont
    \begin{tabular}{cccccccccccc}
        \toprule
        $\rho_2^{12}$ & ($\rho_0^1$, $\rho_1^1$) & ($\rho_0^2$, $\rho_1^2$) & ($\rho_0^{12}$, $\rho_1^{12}$) & ($\delta_1/\sigma_{\delta 1}$, $\delta_2/\sigma_{\delta 2}$) & $I$ & $N$ & $T$ & $(e_1,e_2)$ & $\psi$ & $\bar{\psi}$\\
        \midrule
        0.2 & (0.02, 0.01) & (0.02, 0.01) & (0.01, 0.005) & (0.43, 0.43) & 20 & 13 & 3 & (3.5, 3.0) & 84.5 & 83.0\\
            & & (0.10, 0.05) & (0.01, 0.005) & (0.40, 0.38) & 12 & 25 & 5 & (3.8, 4.6) & 85.2 & 86.7\\
            & & (0.20, 0.10) & (0.01, 0.005) & (0.39, 0.56) & 12 & 25 & 4 & (3.9, 4.3) & 83.6 & 86.2\\
            & (0.10, 0.05) & (0.02, 0.01) & (0.01, 0.005) & (0.38, 0.33) & 12 & 25 & 5 & (3.5, 4.2) & 82.6 & 85.0\\
            & & (0.1, 0.05) & (0.05, 0.025) & (0.49, 0.98) & 12 & 15 & 4 & (4.3, 3.7) & 85.6 & 88.1\\
            & & (0.20, 0.10) & (0.05, 0.025) & (0.59, 0.99) & 12 & 20 & 3 & (4.6, 4.9) & 84.2 & 84.7\\
            & (0.20, 0.10) & (0.02, 0.01) & (0.01, 0.005) & (0.47, 0.22) & 20 & 18 & 5 & (5.5, 4.5) & 82.2 & 82.3\\
            & & (0.10, 0.05) & (0.05, 0.025) & (0.92, 0.92) & 10 & 12 & 3 & (3.8, 3.5) & 84.1 & 84.8\\
            & & (0.20, 0.10) & (0.10, 0.05) & (0.54, 0.81) & 12 & 25 & 4 & (4.9, 3.9) & 83.9 & 85.8\\
        \hline
        0.5 & (0.02, 0.01) & (0.02, 0.01) & (0.01, 0.005) & (0.30, 0.28) & 30 & 10 & 4 & (4.8, 4.7) & 84.4 & 84.1\\
            & & (0.10, 0.05) & (0.01, 0.005) & (0.34, 0.88) & 16 & 22 & 3 & (3.3, 4.2) & 82.4 & 81.3\\
            & & (0.20, 0.10) & (0.01, 0.005) & (0.42, 0.83) & 8 & 20 & 5 & (1.2, 3.3) & 86.3 & 86.2\\
            & (0.10, 0.05) & (0.02, 0.01) & (0.01, 0.005) & (0.38, 0.55) & 21 & 10 & 4 & (4.8, 5.2) & 84.0 & 84.7\\
            & & (0.10, 0.05) & (0.05, 0.025) & (0.52, 0.68) & 8 & 25 & 5 & (3.8, 3.1) & 84.8 & 88.7\\
            & & (0.20, 0.10) & (0.05, 0.025) & (0.62, 0.62) & 22 & 8 & 3 & (5.4, 4.9) & 83.9 & 83.9\\
            & (0.20, 0.10) & (0.02, 0.01) & (0.01, 0.005) & (0.84, 0.29) & 26 & 18 & 3 & (4.9, 4.7) & 84.7 & 86.8\\
            & & (0.10, 0.05) & (0.05, 0.025) & (0.60, 0.60) & 12 & 16 & 4 & (4.7, 5.3) & 85.0 & 85.8\\
            & & (0.20, 0.10) & (0.10, 0.05) & (0.32, 0.84) & 24 & 24 & 5 & (5.1, 4.6) & 85.7 & 86.4\\
        \hline
        0.8 & (0.02, 0.01) & (0.02, 0.01) & (0.01, 0.005) & (0.31, 0.55) & 12 & 16 & 5 & (4.1, 5.0) & 84.4 & 82.6\\
            & & (0.10, 0.05) & (0.01, 0.005) & (0.29, 0.57) & 30 & 14 & 3 & (4.0, 4.7) & 83.1 & 84.6\\
            & & (0.20, 0.10) & (0.01, 0.005) & (0.20, 0.84) & 30 & 17 & 4 & (4.4, 5.4) & 81.4 & 80.2\\
            & (0.10, 0.05) & (0.02, 0.01) & (0.01, 0.005) & (0.31, 0.62) & 20 & 13 & 5 & (5.1, 3.2) & 84.2 & 83.5\\
            & & (0.10, 0.05) & (0.05, 0.025) & (0.82, 0.92) & 8 & 22 & 3 & (2.9, 2.8) & 85.2 & 87.4\\
            & & (0.20, 0.10) & (0.05, 0.025) & (0.45, 0.45) & 18 & 18 & 4 & (5.2, 4.6) & 83.7 & 85.4\\
            & (0.20, 0.10) & (0.02, 0.01) & (0.01, 0.005) & (0.99, 0.25) & 28 & 25 & 3 & (5.1, 4.0) & 85.6 & 84.9\\
            & & (0.10, 0.05) & (0.05, 0.025) & (0.63, 0.31) & 24 & 17 & 4 & (4.5, 5.6) & 84.1 & 84.6\\
            & & (0.20, 0.10) & (0.10, 0.05) & (0.82, 0.82) & 8 & 10 & 5 & (3.2, 2.9) & 86.1 & 89.4\\
        \bottomrule
    \end{tabular}
    }
\end{table}


\section{Discussion}
\label{s:discussion}

Cluster randomized trials with multivariate or co-primary outcomes are becoming increasingly common. Investigators often reluctantly choose a single primary outcome even though there could be multiple outcomes identified by various trial stakeholders as central to their decision-making processes. The recent review of pragmatic Alzheimer’s disease and related dementias trials by Taljaard et al.\citep{taljaard2021methodological} suggested that a substantial proportion of CRTs had multivariate or co-primary outcomes, but appropriate methods for power analysis are not accessible or remain to be developed. Specifically, while methods for designing parallel-arm cluster randomized trials with multivariate outcomes were only recently developed \citep{li2020power,yang2022power}, no such methods were available for designing SW-CRTs with multivariate outcomes. To fill this gap, we developed computationally efficient sample size calculations for designing SW-CRTs with multivariate continuous outcomes using a MLMM. Our model specification includes five ICC parameters representing the endpoint-specific ICCs and between-endpoint ICCs, both within-period and between-period, and the intra-subject ICC. We derive the joint distribution of the intervention test statistics which can be used for generating power estimates under any specified hypothesis and provide an example using the commonly utilized IU-test for co-primary endpoints. We provide insights into the relationship between the ICCs and power. Specifically, we found that higher values of the within-period endpoint-specific ICCs and lower values of all remaining ICCs appear to lead to conservative sample sizes; this may help formulate guidance that is useful when there is limited knowledge regarding the exact value of each ICC (for example, to understand whether larger or smaller ICC values can lead to conservative and therefore still valid sample size estimates in the design phase) but caution that the true relationship between ICCs and power could be more complex. We also conducted an extensive simulation study to validate our power and sample size methodology under small ICCs, small number of clusters, and small cluster-period sizes. Based on these results, we recommend for studies with small expected ICCs (intra-subject ICC as low as 0.2 with all remaining ICCs between 0.005 and 0.02) to have at least 12 clusters with no fewer than 25 participants per cluster-period or at least 20 clusters with no fewer than 15 participants per cluster-period for valid power estimates. However, this guidance only pertains to two co-primary outcomes ($K=2$), which is arguably the most common case in practice; it is possible that this requirement may change under three or more co-primary outcomes and additional investigation is required to explore a more general rule of thumb for $K\geq 3$.
Furthermore, we discuss power calculation under simplified scenarios, including common treatment effects and common ICCs, and under extensions to a more complex closed-cohort design. We illustrate our power formula using the IP-SDM study and assess the sensitivity of our ICC specifications to predicted power. 

Our work assumes an immediate and sustained effect of the intervention on all primary endpoints, and have not considered gradually increasing or decreasing treatment effects by duration of the intervention (referred to as the exposure time). It is of interest to further explore design and analysis strategies for SW-CRTs with multivariate outcomes when the intervention effects are unknown functions of the exposure time. Under the SW-CRT design with a single primary endpoint, Kenny et al.\citep{kenny2021analysis} found that assuming an immediate treatment effect when the true treatment effect is a function of exposure time can lead to estimation bias and invalid inference. It is likely that the same conclusion would apply to the multivariate linear mixed models with more than one primary endpoint. It would also be of interest to develop corresponding power and sample size strategies to detect the existence of exposure-time treatment effect heterogeneity. In addition, we focused on multivariate continuous outcomes, but our method could be extended to binary or other distributions by changing our MLMM to a multivariate generalized LMM (GLMM). The use of a multivariate GLMM is complicated by the requirement of designating an appropriate link function. Based on the power formula for a SW-CRT with a single outcome under the GLMM framework,\citep{davis2021sample} we would expect the covariance matrix of the intervention effect estimators to depend on the period effects and may require complex integration to eliminate the dependence on the random effects. In future work, it is of interest to investigate whether the linearization approach in Davis-Plourde et al.\citep{davis2021sample} can be extended to a multivariate GLMM with multivariate binary outcomes in SW-CRTs, or whether the generalized estimating equations approach for parallel-arm CRTs with multivariate binary outcomes \citep{li2020power} could be extended to more complex correlation structures under a SW-CRT design. Finally, it is also of future interest to develop suitable sample size methodology for SW-CRTs with a mixture of correlated continuous and binary outcomes.




\vspace{0.5cm}
\noindent\textbf{\large{Acknowledgements}}\\
This work is supported by the National Institute of Aging (NIA) of the National Institutes of Health (NIH) under Award Number U54AG063546, which funds NIA Imbedded Pragmatic Alzheimer’s Disease and AD-Related Dementias Clinical Trials Collaboratory (NIA IMPACT Collaboratory). Fan Li is also supported by a Patient-Centered Outcomes Research Institute Award\textsuperscript{\textregistered} (PCORI\textsuperscript{\textregistered} Award ME-2020C3-21072). The content is solely the responsibility of the authors and does not necessarily represent the official views of the NIH, PCORI\textsuperscript{\textregistered} or its Board of Governors or Methodology Committee. The authors are grateful to Lionel Adisso and Dr. France Légaré for providing data and information from the IP-SDM study.

\vspace{0.5cm}
\noindent\textbf{\large{Data Availability Statement}}\\
The data used to generate parameter estimates for illustration of sample size methodology were obtained from the IP-SDM investigators. Restrictions may apply to the availability of these data. Data requests can be submitted to the corresponding author (kendra.plourde@yale.edu) who will correspond with the IP-SDM investigators to obtain data permissions.

\bibliography{WileyNJD-AMA}%

\vspace{1cm}
\noindent\textbf{\large{Supporting Information}}\\
Web Appendices may be found online in the Supporting Information section at the end of this article. R code for predicting power and for conducting the simulation and application studies are available at \url{https://github.com/kldavisplourde/SWCRTmultivariate}.

\end{document}